%% file: main.tex
\documentclass[twoside,twocolumn,9pt]{article}
\usepackage{extsizes}
\usepackage[super,sort&compress,comma]{natbib} 
\usepackage[version=3]{mhchem}
\usepackage[left=1.5cm, right=1.5cm, top=1.785cm, bottom=2.0cm]{geometry}
\usepackage{amsmath}
\usepackage{balance}
\usepackage{mathptmx}
\usepackage{sectsty}
\usepackage{graphicx} 
\usepackage{lastpage}
\usepackage[format=plain,justification=justified,singlelinecheck=false,font={stretch=1.125,small},labelfont=bf,labelsep=space]{caption}
\usepackage{float}
\usepackage{subcaption}
\usepackage{fancyhdr}
\usepackage{fnpos}
\usepackage[english]{babel}
\addto{\captionsenglish}{%
  
}
\usepackage{array}
\usepackage{siunitx}
\usepackage{droidsans}
\usepackage{charter}
\usepackage[T1]{fontenc}
\usepackage{color}
\usepackage[usenames,dvipsnames]{xcolor}
\usepackage{setspace}
\usepackage[compact]{titlesec}
\usepackage{comment}
\usepackage[inline]{enumitem}
\usepackage{multirow}
\usepackage{adjustbox}
\usepackage{tikz}
\usepackage{authblk}
\usepackage{hyperref}

\graphicspath{ {Images/} }

\hypersetup{
    colorlinks,
    linkcolor={red!50!black},
    citecolor={blue!50!black},
    urlcolor={blue!80!black}
}

\usetikzlibrary{arrows.meta,positioning}
\tikzset{every node/.style={inner sep=0,outer sep=0}}
\tikzset{>=Triangle}

\newcommand{\fig}[1]{Figure~\ref{#1}}
\newcommand{\cm}{\si{cm^{-1}}}

\newcommand{\sisio}{\si{Si/SiO_2}}
\newcommand{\meet}{4:1~Me\-tha\-nol\-:\-E\-tha\-nol}

\newcommand{\scalexfifty}[1]{\draw [white, line width = #1/40] (#1*.95/2,-#1*.35) -- node [midway, above=#1/30] {\textsf{\textbf{\scriptsize 40 µm}}} +(-#1*0.309,0);}

\definecolor{Czero}{rgb}{0.12156, 0.46666, 0.7058}
\definecolor{Cone}{rgb}{1, .498, .055}
\definecolor{cream}{RGB}{222,217,201}

\title{Strain and doping transfer between suspended and supported bilayer graphene }
\author[1]{Riccardo Galafassi}
\author[1]{Fabien Vialla}
\author[1]{V. Rajaji}
\author[1] {Alexis Forestier}
\author[2]{Bruno Sousa Araújo}
\author[1]{Hatem Diaf}
\author[1,3]{Natalia Del Fatti}
\author[2]{Antonio Gomes Souza Filho}
\author[4] {Arnaud Claudel}
\author[4] {La\"etitia Marty}
\author[1]{Alfonso San-Miguel}

\affil[1] {Université Claude Bernard Lyon 1, CNRS, Institut Lumière Matière UMR 5306, F-69622, Villeurbanne,	France}
\affil[2] {Departamento de Física, Centro de Ciencias,Universidade Federal Do Ceara, CEP 60.455-970, Fortaleza, Brazil}
\affil[3] {Institut Universitaire de France (IUF), Paris, France}
\affil[4] {Université Grenoble Alpes, CNRS, Grenoble INP, Institut NEEL, 38000, Grenoble, France}

\date{}

\begin{document}

\pagestyle{fancy}
\thispagestyle{plain}
\fancypagestyle{plain}{
\renewcommand{\headrulewidth}{0pt}
}

\makeFNbottom
\makeatletter
\renewcommand\LARGE{\@setfontsize\LARGE{15pt}{17}}
\renewcommand\Large{\@setfontsize\Large{12pt}{14}}
\renewcommand\large{\@setfontsize\large{10pt}{12}}
\renewcommand\footnotesize{\@setfontsize\footnotesize{7pt}{10}}
\makeatother

\renewcommand{\thefootnote}{\fnsymbol{footnote}}
\renewcommand\footnoterule{\vspace*{1pt}%
\color{cream}\hrule width 3.5in height 0.4pt \color{black}\vspace*{5pt}} 
\setcounter{secnumdepth}{5}

\makeatletter 
\renewcommand\@biblabel[1]{#1}            
\renewcommand\@makefntext[1]%
{\noindent\makebox[0pt][r]{\@thefnmark\,}#1}
\makeatother 
\renewcommand{\figurename}{\small{Fig.}~}

\subsectionfont{\normalsize}
\subsubsectionfont{\bf}
\setstretch{1.125} 
\setlength{\skip\footins}{0.8cm}
\setlength{\footnotesep}{0.25cm}
\setlength{\jot}{10pt}
\titlespacing*{\section}{0pt}{4pt}{4pt}
\titlespacing*{\subsection}{0pt}{15pt}{1pt}

\fancyhead{}
\renewcommand{\headrulewidth}{0pt} 
\renewcommand{\footrulewidth}{0pt}
\setlength{\arrayrulewidth}{1pt}
\setlength{\columnsep}{6.5mm}
\setlength\bibsep{1pt}

\makeatletter 
\newlength{\figrulesep} 
\setlength{\figrulesep}{0.5\textfloatsep} 

\newcommand{\topfigrule}{\vspace*{-1pt}%
\noindent{\color{cream}\rule[-\figrulesep]{\columnwidth}{1.5pt}} }

\newcommand{\botfigrule}{\vspace*{-2pt}%
\noindent{\color{cream}\rule[\figrulesep]{\columnwidth}{1.5pt}} }

\newcommand{\dblfigrule}{\vspace*{-1pt}%
\noindent{\color{cream}\rule[-\figrulesep]{\textwidth}{1.5pt}} }

\makeatother

\maketitle
\begin{abstract}
\input{Chapters/Abstract}
\end{abstract}

\section{Introduction}
\input{Chapters/Introduction}

\section{Experimental methods}
\input{Chapters/Experimental}
\section{Results and discussion}\label{Sec:Results}
\input{Chapters/Results}

\section{Conclusion}
\input{Chapters/Conclusion}

\section*{Declaration of competing interest}
The authors declare that they have no known competing ﬁnancial interests or personal relationships that could have appeared to inﬂuence the work reported in this paper.

\section*{Data availability statement}
The data supporting this study are available from the corresponding author upon reasonable request.

\section*{Acknowledgements}
The authors acknowledge the support from ANR (project 2D-PRESTO ANR-19-CE09-0027).

\bibliography{refs} 
\bibliographystyle{rsc}
\end{document}

%% file: Chapters/Abstract.tex
Due to their unique dimensionality, the physical properties of two-dimensional materials are deeply impacted by their surroundings, calling for a thorough understanding and control of these effects. 
We investigated the influence of the substrate and the pressure transmitting medium on bilayer graphene in a unique high-pressure environment where the sample is partially suspended and partially supported. 
By employing Raman spectroscopy with a sub-micron spatial resolution, we explored the evolution of strain and doping, and demonstrated that they are both similarly induced in the suspended and supported regions of the bilayer graphene within the studied pressure range. 
Almost full strain and doping transfer between the supported and suspended regions is concluded. 
We observed that charge carrier density saturates quickly at low pressures (2 GPa) while biaxial strain continuously increases with pressure.
Additionally, Raman spatial mapping highlights a rather uniform doping and strain distribution, yet with significant local variations revealing a more complex scenario than previously documented by single-point studies at high pressure.

%% file: Chapters/Introduction.tex
In the era of rapid scientific advancements, two-dimensional (2D) materials have emerged as a captivating field of study, revolutionizing our understanding of fundamental physics, enabling novel device functionalities and promising breakthroughs in various technological applications. Their success resides in their outstanding electronic\cite{Novoselov2004,Zhang2005}, optical\cite{Yang2016,Feng2008} and thermal\cite{Balandin2008} properties derived from their 2D nature. This thrilling mix is complemented by their nanometric thickness resulting in extremely light materials, while preserving exceptional mechanical properties\cite{Changgu2008,Sun2021}. 
Extensive studies have also been directed into the possibility of tuning 2D materials properties through external variables. Temperature\cite{Bolotin2008,Taube2015}, magnetic\cite{Burch2018,Mak2019} and electric fields\cite{Zhang2009,Jautegui2019,Chiout2023} are commonly used whilst pressure is often overlooked. The latter can act on those materials in various ways. With pressure application, the electronic structure can be extensively altered leading to modification of the optical properties\cite{Medeghini2018,Machon2018b}, with closing and opening of the band gap\cite{Martins2017}. When critical densities are reached phase transitions occur, drastically modifying the materials properties\cite{Martins2017,vincent2024}. 

While those considerations are quite general and may also apply to bulk materials, when looking to 2D systems, further effects can result from their reduced dimensionality. By construction, the majority or totality of their atoms constitute surface atoms. They are, thus, extremely sensitive to the surrounding environment. Pressure affects the environment increasing the electronic density and amplifying the interactions between the 2D system and its local environment. Environmental effects on 2D materials in high pressure experiments are thus thoroughly studied focusing on two main elements: the substrate and the so called pressure transmitting medium (PTM). The substrate is a fundamental element of 2D materials providing their support. Additionally, it is a source of mechanical strain\cite{Bronsgeest2015,Schmidt2011} and doping\cite{Fan2011,JI2021,Sun2022}.
The second element, the PTM, is an essential component of high pressure experiments as it is used to transfer pressure from a compression system to the sample. It therefore surrounds the studied sample and interacts with it\cite{Proctor2006,Machon2018,Forestier2020}.  
High-pressure experiments in 2D materials allow for highly amplified interactions between the 2D system and both the substrate and the PTM, in contrast to bulk materials. Due to these intensified interactions, entirely new physical phenomena can emerge. It appears then fundamental to quantify or separate the effect of each contribution.

Driven by these principles we focus in this work on investigating the effect of the two mentioned environmental elements, the substrate and the PTM, on the pressure response of 2D materials. As a benchmark material we choose graphene as it has been shown to strongly interact with these elements when compressed \cite{Bousige2017,Machon2018,Forestier2020, vincent2024}. In particular, major attention was given to the investigation of their strain and doping contributions in compressed graphene. In order to achieve this, we introduced a novel configuration for high pressure experiments, in which the same sample is found both suspended and supported on a substrate. This binary system opens up the opportunity to study, within the same pressure run, the same sample subjected to different environmental conditions and allows for the disentanglement of substrate and PTM contributions in the evolution of graphene’s features. To probe the graphene response at high pressure we used Raman spectroscopy. This technique is widely exploited for measuring doping and strain in graphene both at ambient conditions and at high pressure\cite{Machon2018,Forestier2020,Lee2012,Bousige2017}.

%% file: Chapters/Experimental.tex
Our experiments were conducted focusing on the investigation of the local high pressure response of graphene in the chosen suspended/supported configuration. Thus, we performed our measurements using a novel cell design developed in our laboratory, designed to achieve a high degree of spatial resolution~\cite{Medeghini2018}. In conventional high pressure experiments that uses diamond anvil cells (DACs) the aberrations introduced by the diamond interposed between the sample and the spectrometer severely worsen the optical resolution. Moreover, the bulky nature of conventional DACs, does not allow for using high magnification microscope objectives with an instrumental limitation set to long-working-distance x50 objectives. Our cell was specifically designed to allow the use of a modified Mitutoyo x100 magnification objective with 0.7~NA. This objective was conceived with a correction for spherical aberrations introduced by the passage of light through the diamond in the optical path. This set-up allows to achieve sub-micron resolution in high pressure experiments~\cite{Medeghini2018}. This allowed to ensure that we investigated the contributions from the suspended and supported regions of graphene individually. Moreover, we performed detailed high pressure Raman spectroscopy measurements and spatial mapping of our sample.

\begin{figure*}
\centering 
\newcommand{\aim}[2]{
    \node (c) at (#1) {};
	\begin{scope}[scale=0.2]
	    \fill  [#2, line width=.7, even odd rule, draw=black] (c) circle (1) (c) circle (.6);
	    \draw [line width=.1] (c) +(.15,0) -- +(-.15,0)
	    	 +(0,.15) -- +(0,-.15);
	    \foreach \x in {0,...,3}
	        \draw [line width=.8] (c) ++(90*\x:.3) -- +(90*\x:1.3);
	\end{scope}
}

\begin{subfigure}{.33\textwidth}
		\begin{tikzpicture}
		\node () at (0,0) {\includegraphics[width = \textwidth]{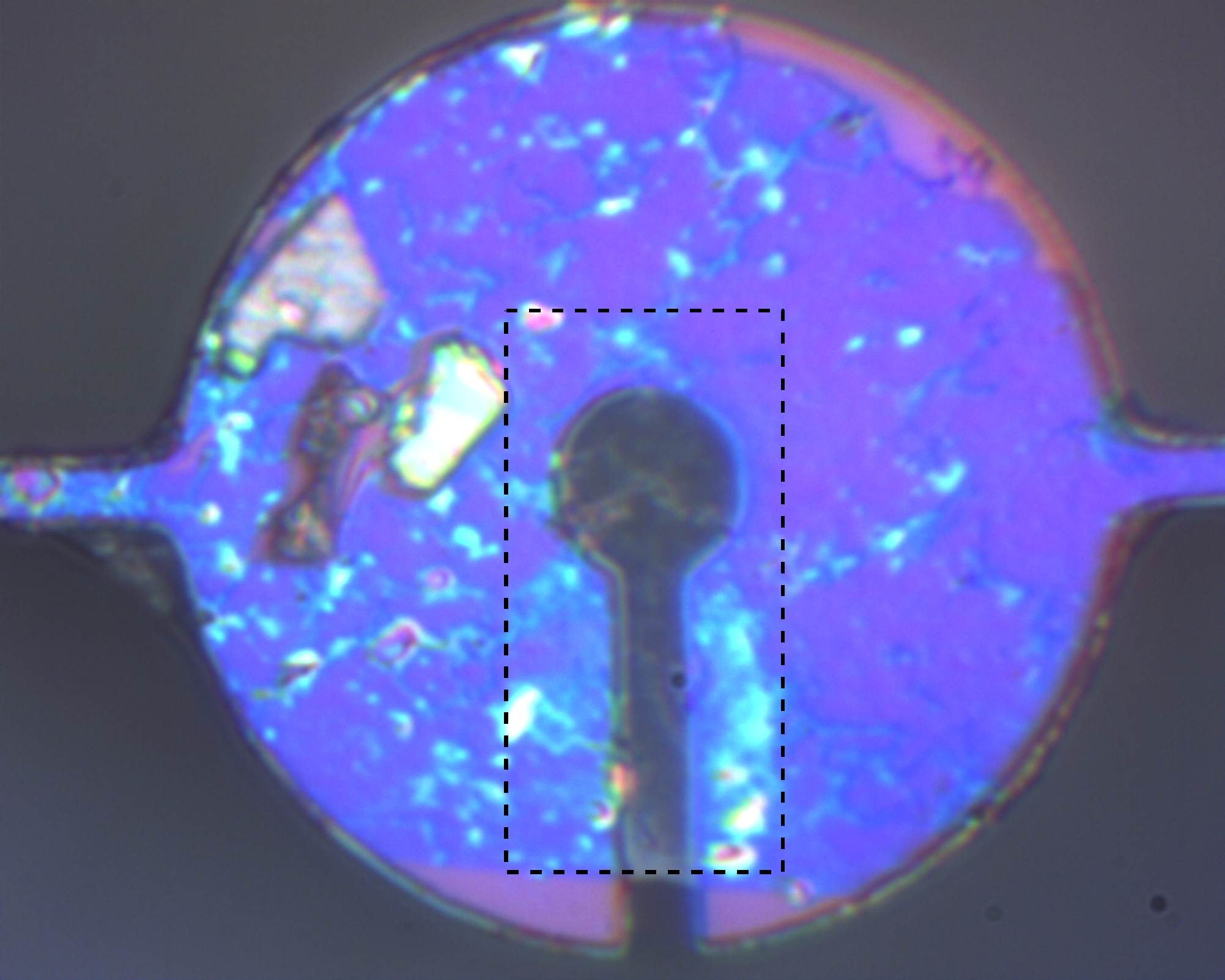}};
		\aim {0.2,0.1}{Czero};
		\aim {1.4,0.1}{Cone};
		\scalexfifty{\textwidth}
		\end{tikzpicture}
		\caption{Ambient}\label{Fig:sample_P00}
\end{subfigure}
~~~~
\begin{subfigure}{.33\textwidth}
		\begin{tikzpicture}
		\node () at (0,0) {\includegraphics[width = \textwidth, angle = 180]{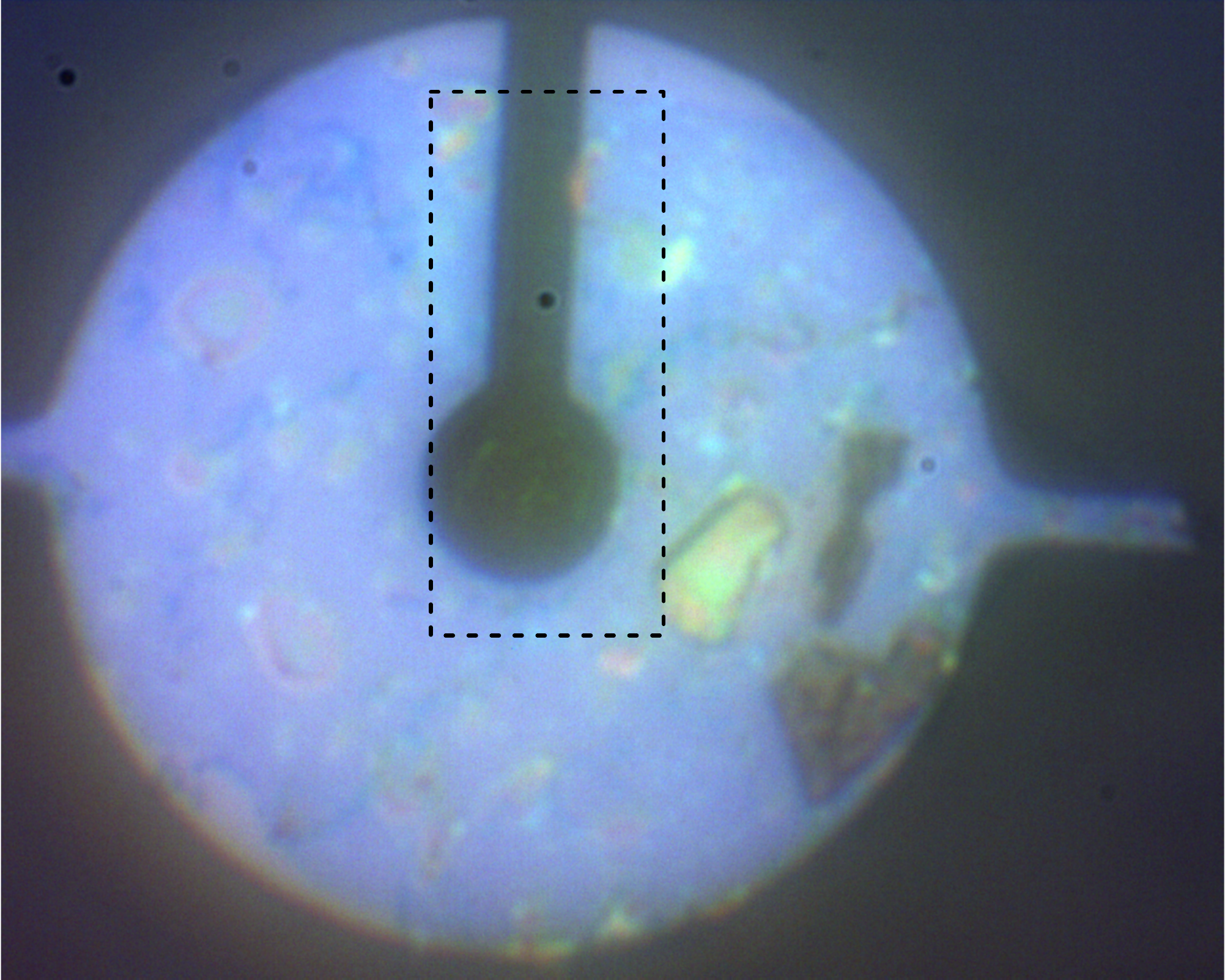}};
		\draw [<-, red] (0.03\textwidth,-.33\textwidth) -- +(200:.15\textwidth);
		\scalexfifty{\textwidth}
		\end{tikzpicture}
		\caption{0.6~GPa}\label{Fig:sample_P02}
\end{subfigure}

\begin{subfigure}{.5\textwidth}
		\includegraphics[width = \textwidth]{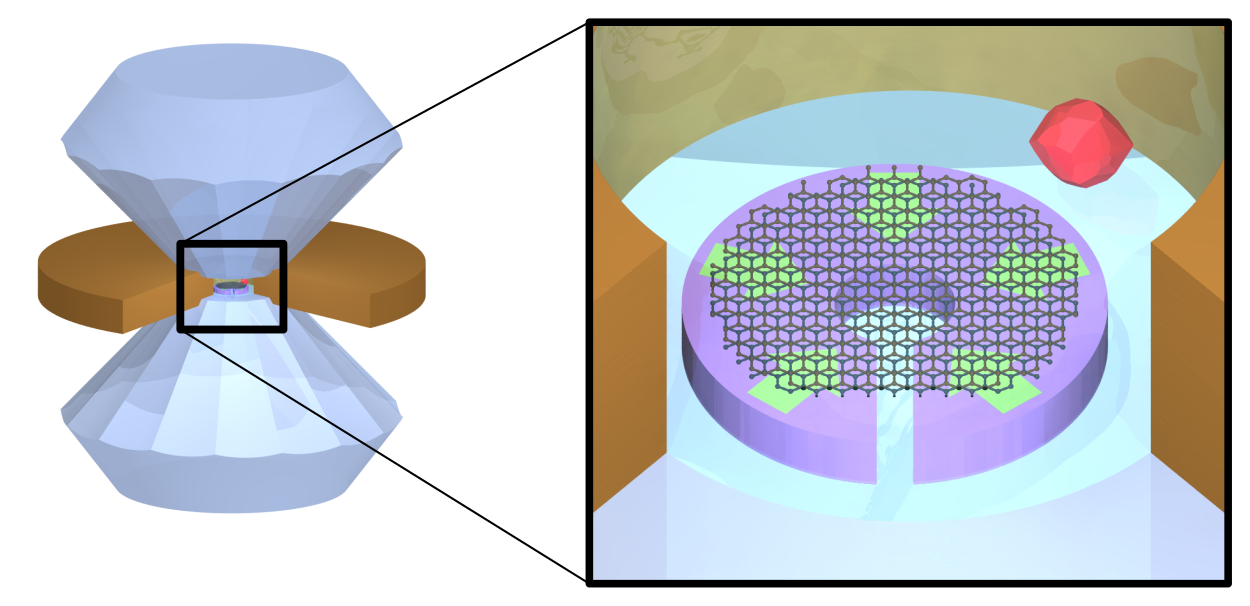}
\caption{}\label{Fig:DAC}
\end{subfigure}

\begin{subfigure}{.7\textwidth}
		\centering
		\includegraphics[width = \textwidth]{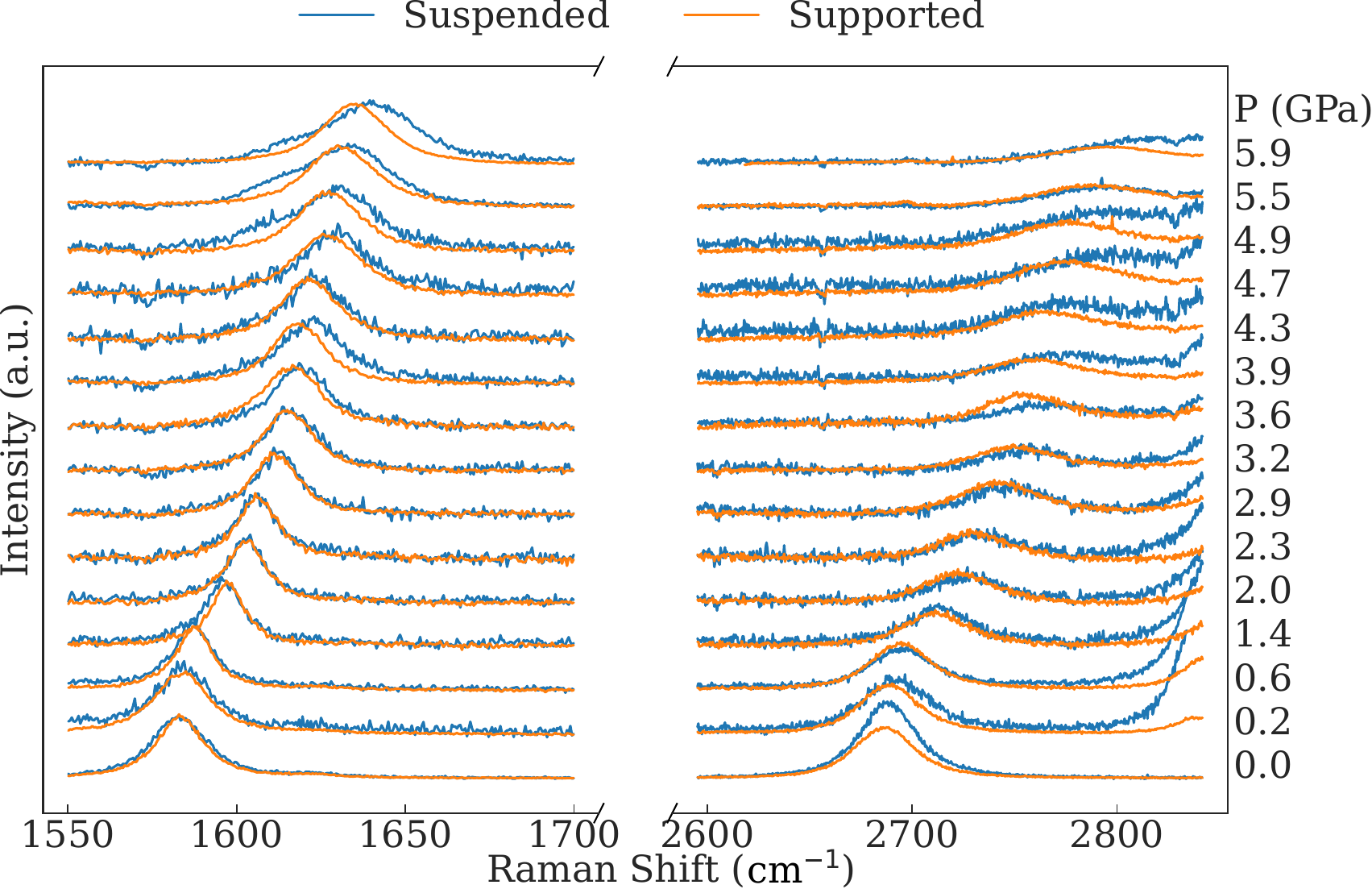}
\caption{}\label{Fig:2D:Stack}
	\end{subfigure}

\caption{Optical image at x50 magnification of the sample before (\subref{Fig:sample_P00}) and after (\subref{Fig:sample_P02}) loading in the DAC. A black dashed rectangle indicates the region that has been mapped through Raman spectroscopy. In (\subref{Fig:sample_P00}) an orange and a blue sights show the points measured throughout the pressure cycle, later referred as Supported and Suspended respectively. A red arrow in (\subref{Fig:sample_P02}) points to the BLG edge to indicate a faint contrast that confirms the presence of the sample in the suspended region after loading the cell with the PTM. (\subref{Fig:DAC}) Schematic representation of the DAC system used in our experiment. The two anvils (in light blue) compress the gasket (in brown). In the black rectangle, a zoomed-in view of the sample area shows a red ruby chip and the BLG sample on the substrate. Green arrows on the substrate schematically indicates the biaxial strain applied to the BLG due to the substrate’s compression under pressure. (\subref{Fig:2D:Stack}) Raman spectra of the suspended and supported BLG collected in the high pressure run. The two regions correspond to the G-band (below 1700~\cm{}) and the 2D-band (above 2600~\cm{}). The spectra are normalized to the G-band maximum intensity for each pressure.}\label{Fig:Sample}
\end{figure*}

The experiments were conducted using a bilayer graphene (BLG) sample. Initial tests with monolayer graphene led to the breakage of the sample in the suspended region when subjected to the high pressure environment. BLG, on the other hand, showed a higher success rate in resisting the cell loading procedure, so it was chosen for our experiments. The BLG sample we studied was supported by a small circular disc with a diameter of $\sim 90$~\si{\micro\meter} etched out using lithography from a 50~\si{\micro\meter} thin \sisio{} foil, compatible with the DAC sample area. A circular through hole of $\sim20$~\si{\micro\meter} was drilled at the centre of the disc in order to accommodate the suspended region of the sample. The dimensions of the hole were initially chosen in an attempt to spatially isolate the suspended and supported regions of the sample and minimize border effects. Moreover, a large hole compared to probing laser spot size would ensure the independent investigation of the two regions. A linear channel was also carved in the disk to allow the PTM to fill the volume below the suspended part of the sample which otherwise would not fill (most likely due to the glue attaching the substrate to the anvil and sealing the access to the hole). Bilayer graphene was deposited on the substrate by two independent wet transfers of CVD-grown monolayer graphene followed by carbon dioxide critical point drying using a Tousimis Autosamdri®-815, Series B. Finally, the sample was deposited on the bottom anvil of our DAC (details of the transfer procedure are given in Section~S1 of the supplementary information (SI)). In \fig{Fig:sample_P00} and \fig{Fig:sample_P02} we can see the sample on the anvil before and after PTM loading respectively.

The choice of the PTM for our experiment is a critical experimental aspect. We found that the alcohol mixture 4:1 Methanol:Ethanol was the most adapted for our purpose. As well as featuring outstanding hydrostaticity up to its solidification around 10.5~GPa\cite{Klotz2009}, it induces doping effects on graphene when compressed\cite{Nicolle2011,Forestier2020}. This allowed us to probe its chemical and physical interactions with graphene, without introducing discontinuities in the PTM’s behaviour, such as phase transitions, that may compromise the interpretation of the results. A comparison between different PTMs was attempted. In particular, the use of water has been shown to produce interesting functionalization effects on graphene, leading to PTM-induced phase transitions\cite{Martins2017,vincent2024}. Furthermore, the use of gaseous PTMs such as argon or nitrogen has shown to mainly induce strain effects on graphene, without affecting the charge doping distribution\cite{Forestier2020}. However, from an experimental point of view, we could not successfully prepare an experiment using other PTM than the alcohol mixture. All the other PTMs led to the breakage of the suspended region of the BLG.
 
A schematic representation of the DAC and the sample area is shown in Figure 1c. A pair of 450~\si{\micro m} cullet size diamonds was used as anvils. A 200~\si{\micro m} thick steel T301 gasket was interposed between the anvils and a 150~\si{\micro m} through hole was drilled at its centre to accommodate the sample compression chamber. In the latter, ruby chips of few micron in diameter were inserted to measure \textit{in-situ} the chamber pressure\cite{Chijioke2005}.

%% file: Chapters/Results.tex
\begin{figure*}
\centering
\includegraphics[width=.6\textwidth]{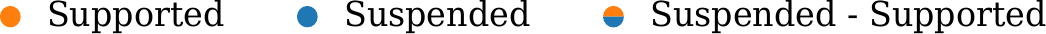}

\begin{subfigure}{.49\textwidth}
\resizebox{\textwidth}{!}{%
	\begin{tikzpicture}
	  	\pgfmathsetmacro{\ratio}{2.9/8*10}
		\node () at (0,0) {\includegraphics[width = 10cm]{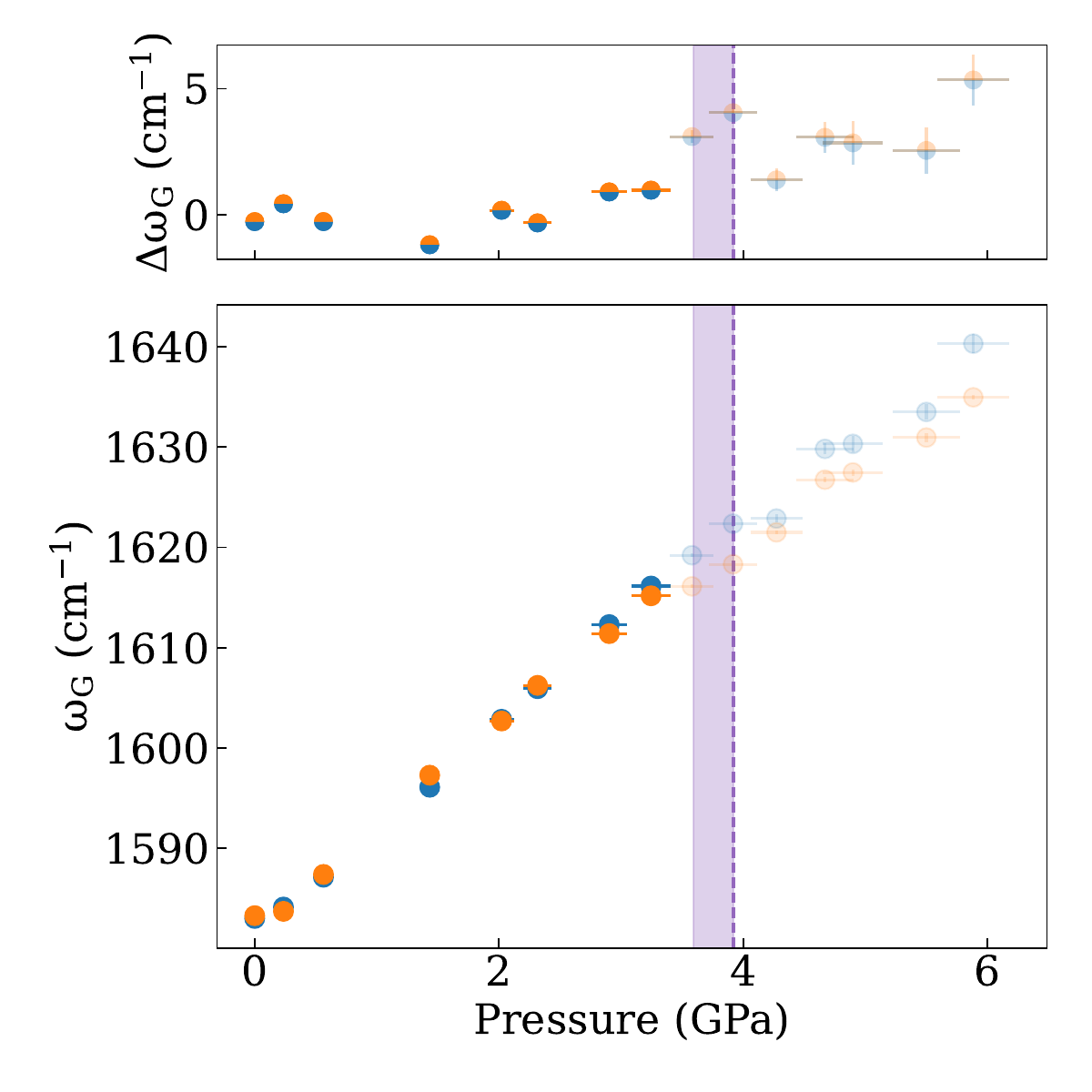}};
	\end{tikzpicture}
}
\caption{}\label{Fig:2D:G_pos}
\end{subfigure}
\begin{subfigure}{.49\textwidth}
\resizebox{\textwidth}{!}{%
	\begin{tikzpicture}
	  	\pgfmathsetmacro{\ratio}{2.9/8*10}
		\node () at (0,0) {\includegraphics[width = 10cm]{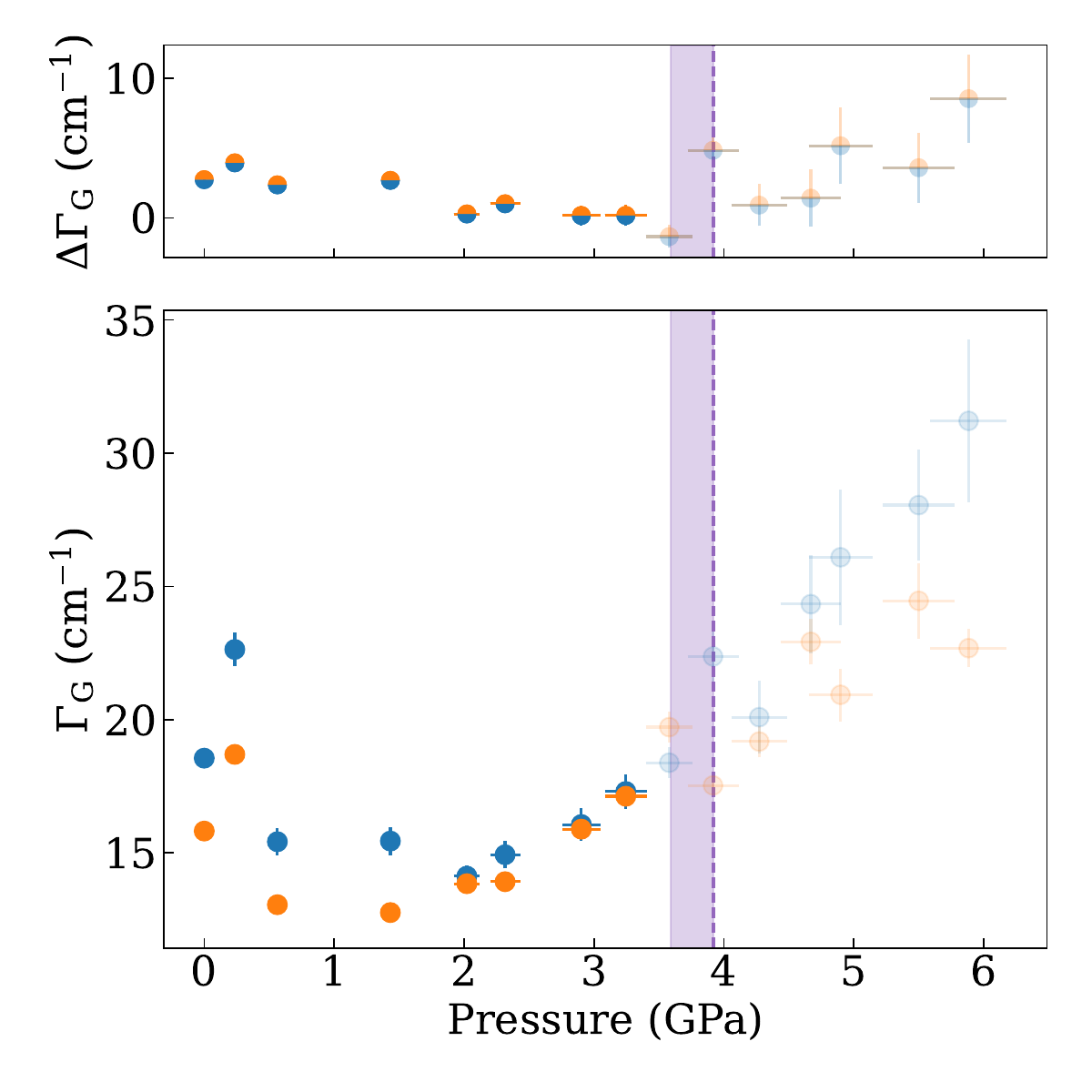}};
		\node () at (0,0) {\includegraphics[width = 10cm]{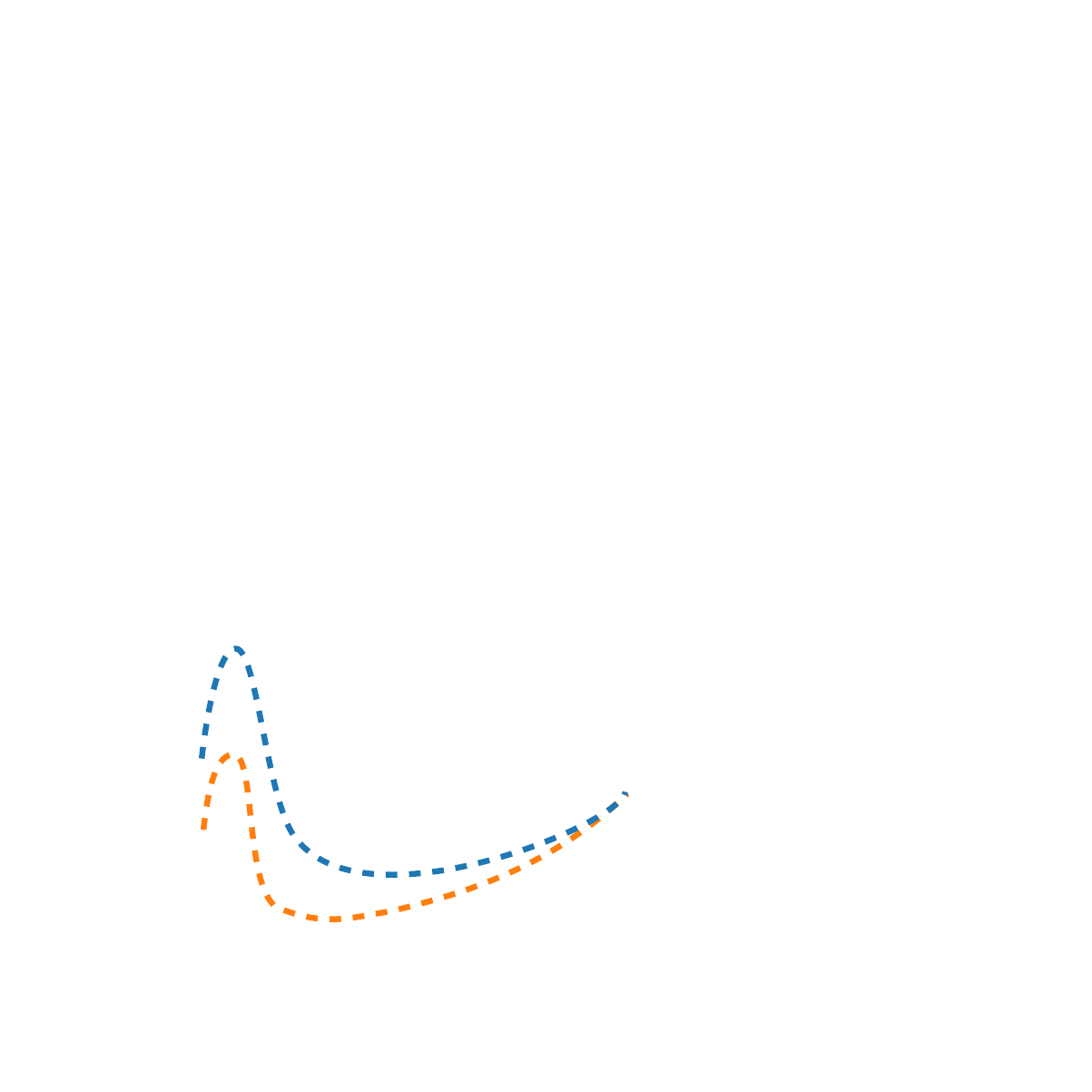}}; 
	\end{tikzpicture}
}
\caption{}\label{Fig:2D:G_width}
\end{subfigure}

\begin{subfigure}{.49\textwidth}
\resizebox{\textwidth}{!}{%
	\begin{tikzpicture}
	  	\pgfmathsetmacro{\ratio}{3.1/8*10}
		\node () at (0,0) {\includegraphics[width = 10cm]{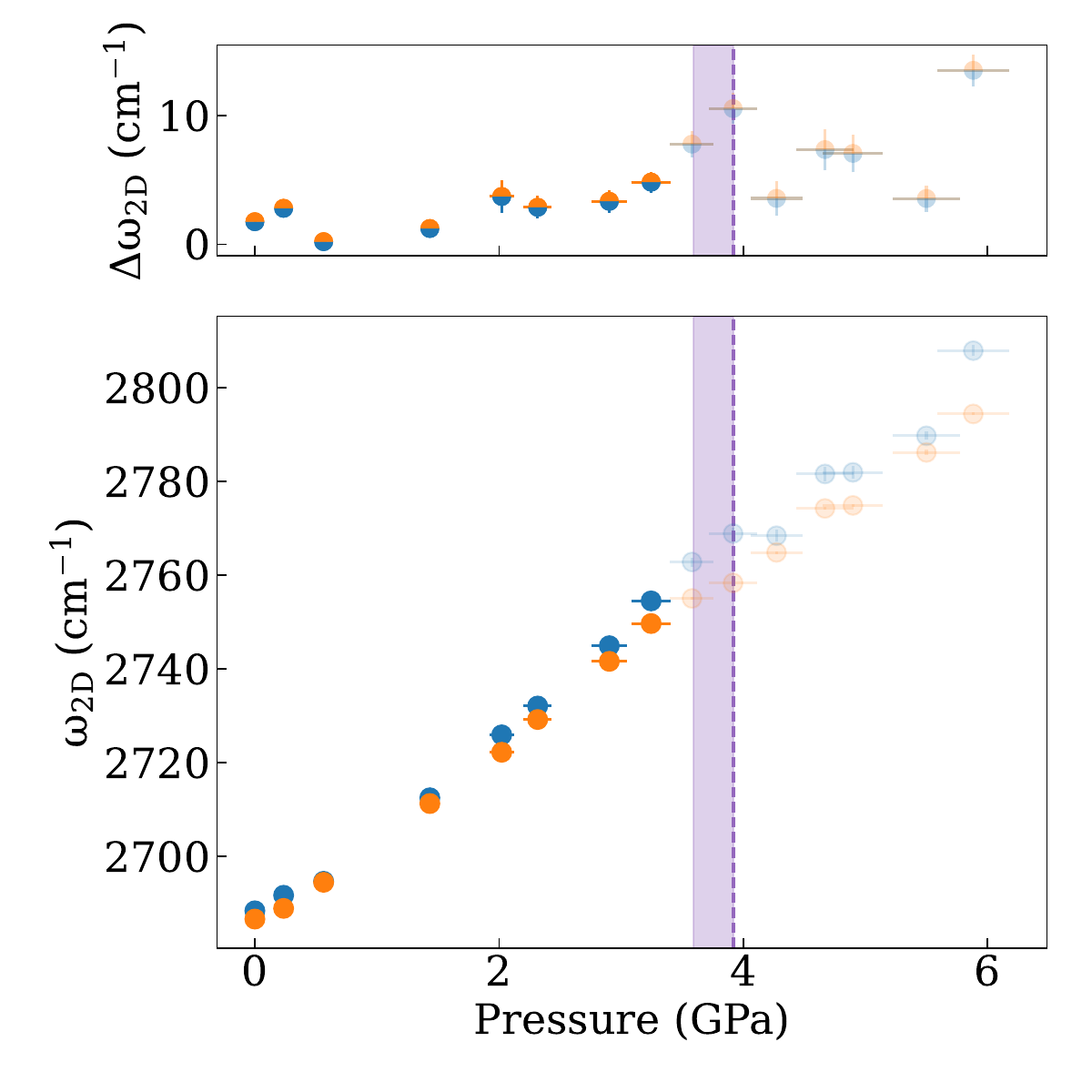}}; 
	\end{tikzpicture}
}
\caption{}\label{Fig:2D:2D_pos}
\end{subfigure}
\begin{subfigure}{.49\textwidth}
\resizebox{\textwidth}{!}{%
	\begin{tikzpicture}
	  	\pgfmathsetmacro{\ratio}{2.9/8*10}
		\node () at (0,0) {\includegraphics[width = 10cm]{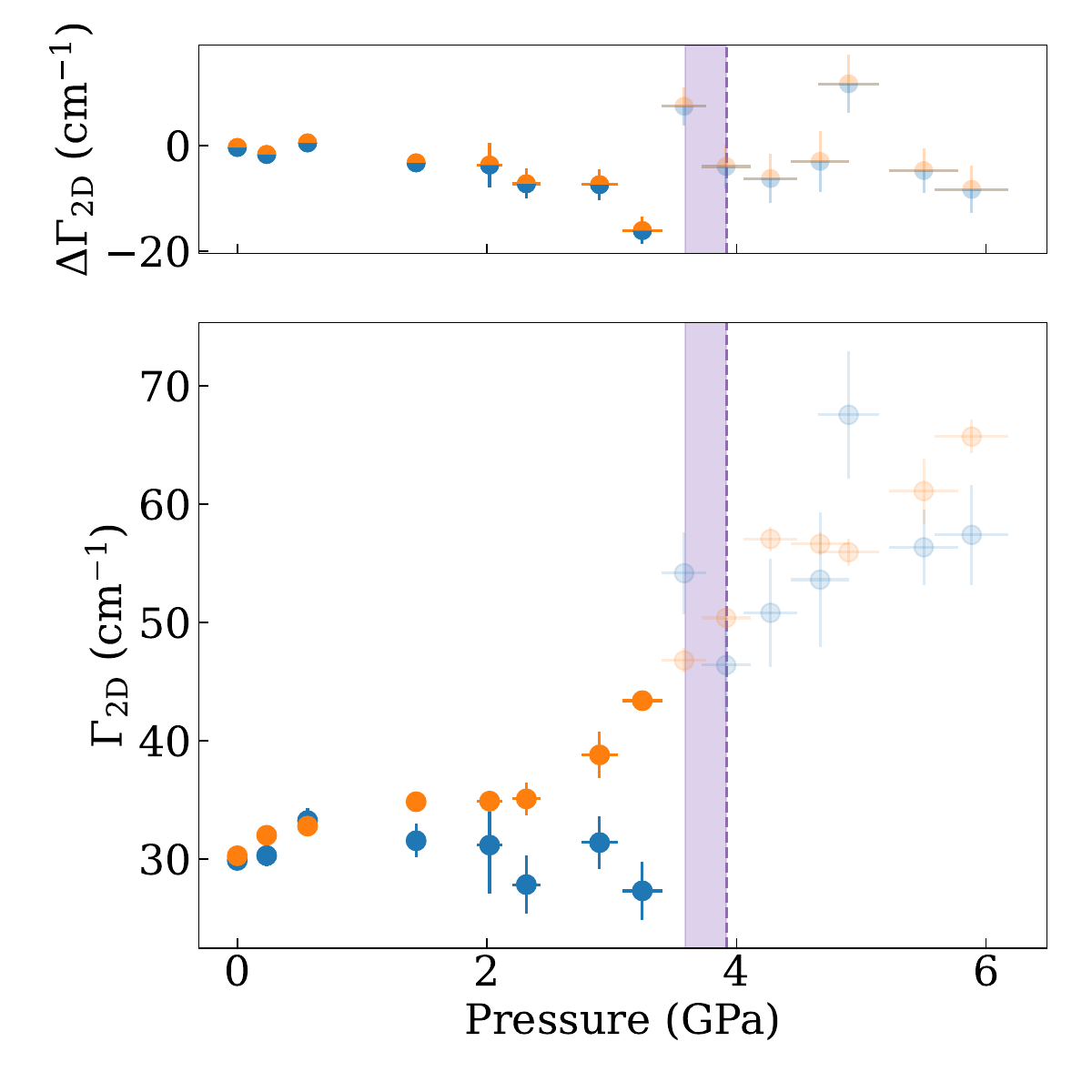}};
	\end{tikzpicture}
}
\caption{}\label{Fig:2D:2D_width}
\end{subfigure}

\caption{Pressure evolution of the fitted G and 2D bands parameters for the spectra in \fig{Fig:2D:Stack}. A purple rectangle indicates the beginning of the delamination of the oxide layer of the substrate. In correspondence of the purple dashed line the transition is observed also by optical imaging. The pressure points above this pressure have been faded as they have not been taken into consideration in our analysis but they have been reported for completeness. Each plot shows the measured value of the parameters (bottom side) and the difference between the corresponding parameter measured in the suspended and supported regions (upper side). In \fig{Fig:2D:G_width} dashed lines have been added as a guide to the eye to help the visualization of the evolution of $\Gamma_G$.}\label{Fig:2D:parameters}
\end{figure*}

We performed a pressure run up to 5.9~GPa monitoring the evolution of the Raman features of the BLG sample. Above this pressure the top diamond anvil touched the sample so we stopped the experiment. The evolution of the G and 2D bands with pressure is shown in \fig{Fig:2D:Stack}. A similar trend is observed in the evolution of the spectra in the two regions at low pressure with a larger blue shift of the suspended features at higher pressure. We observe a neat change in trend around 3.9~GPa, where the width and position of the spectra differ between the suspended and supported regions. At this pressure the delamination of the \si{SiO_2} layer at the substrate surface in contact with the sample is observed by optical imaging (see Figure~S2 in SI). This phenomenon has been previously shown by other works\cite{AMARAL2021,Forestier2020} and it has been attributed	 to the difference of bulk modulus of silicon and silicon oxide composing the substrate.

The spectra have been fitted using Lorentzian functions and the extracted parameters are plotted in \fig{Fig:2D:parameters}. A purple rectangle highlights the pressure at which the \si{SiO_2} delamination is observable by optical means. However, we also note that the previous pressure point at 3.6~GPa largely diverges from the trend at low pressure. This makes us conclude that the delamination might have started earlier but could not be clearly observed in the optical images. The evolution of the sample appearance throughout the pressure cycle can be observed in Figure~S2. The points above the delamination pressure have been faded as the interpretation of the experimental results above this transition is beyond the scope of this work. They will not be considered in the following analysis but they have been kept for completeness. In the following we will refer to $\omega_{i}$ and $\Gamma_{i}$ ($i = G,2D$) as the frequency and the FWHM of the bands, and to $\Delta\omega_{i}$ and $\Delta\Gamma_{i}$ ($i = G,2D$) as the difference between the respective parameters measured in the suspended and supported regions. 

We observe a similar pressure evolution of the frequencies both for the G-band (\fig{Fig:2D:G_pos}) and the 2D-band (\fig{Fig:2D:2D_pos}). A larger blueshift is observed in the case of $\omega_{2D}$ in the suspended region above 2~GPa. $\Delta\omega_G$ and $\Delta\omega_{2D}$ help the visualization of the small variations between the two regions which is more marked in the case of $\omega_{2D}$. This first result is quite striking as intuitively we would expect a larger blueshift in the supported graphene. In fact, the major contribution of the G-band and 2D-band shift with pressure is introduced by a biaxial strain originated from the compression of the substrate\cite{Forestier2020,Machon2018,Bousige2017}. This result indicate that the strain is efficiently transferred from the supported to the suspended part of the sample. Beside strain, when using \meet{} as pressure transmitting medium, previous groups reported an important doping effect of graphene upon compression\cite{Forestier2020,Nicolle2011}. In our case this is also observed by the strong decrease of $\Gamma_G$ at low pressure. While the value of $\Gamma_G$ is lower in the supported region, $\Delta\Gamma_G$ decreases with pressure. This indicates that, relatively, the spectral width of the G-band gets thinner at a faster rate in the suspended region than in the supported one with increasing pressure. This effect could be explained by a larger amount of charges injected in the suspended region due to the presence of the PTM on both sides of the sample. In addition, a local inhomogeneous strain field could be present in the supported graphene due to the substrate's roughness. The observed G-band is thus averaged over the probing laser spot size resulting in a widening of the G-band width which counters the doping effect. This second option is backed up by the observed increase of $\Gamma_G$ above 2~GPa which is not explainable with doping as we would expect a flat evolution after charge saturation\cite{Froehlicher2015}.

When we look at the 2D band width in \fig{Fig:2D:2D_width} we observe a different trend in the two regions. The 2D-band width is fairly constant before delamination for the suspended part of the sample. This result is compatible with the observations of Froelicher \textit{et al.} in electrochemically gated samples where the width of the 2D band is found to be scarcely affected by doping\cite{Froehlicher2015}. On the other hand, the supported part of the BLG features an increase of $\Gamma_{2D}$, which is more marked above 2.3~GPa in agreement with the hypothesis of a inhomogeneous strain field introduced by the substrate. This effect is marked also by the decrease of $\Delta\Gamma_{2D}$ with increasing pressure. In reality, the simultaneous contribution of both effects is the most probable scenario.

Finally, the close observation of $\Gamma_G$ in \fig{Fig:2D:G_width} features an initial widening at 0.2~GPa when compared to the ambient pressure point (acquired without PTM). We can explain this peculiar signature by the presence of an opposite charge doping of graphene at ambient conditions which is neutralized when the PTM is introduced and pressure is increased. Several groups have reported the presence of partial doping of graphene at ambient pressure principally due to impurities on the substrate as well as to environmental pollution\cite{Aguirre2009,Fan2011,JI2021,Wilder1998}. Those groups report a predominant p-doping effect induced by the substrate and the environment on the sample. We thus conclude that a n-doping effect of the PTM is the most probable scenario.

In order to clearly discern the doping and strain contributions of graphene's Raman features we performed a similar procedure to that followed by Lee \textit{et al.}\cite{Lee2012} which will make the object of the following section.

\subsection{Disentangling strain and doping contributions}\label{Sec:Disentangle}

\begin{figure*}
\centering
\includegraphics[width = .8\textwidth]{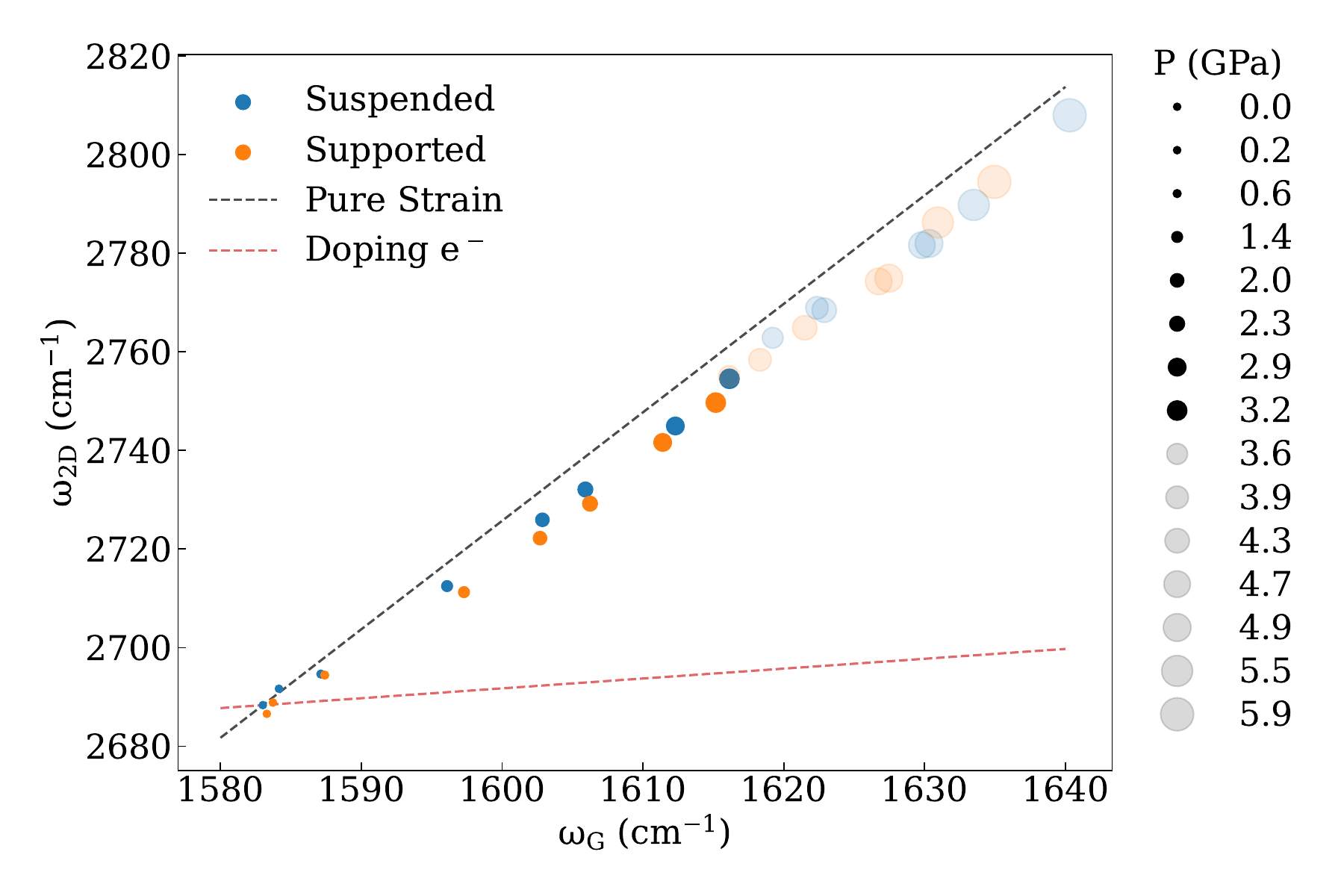}
\caption{Correlation plot for the 2D-band frequency $\omega_{2D}$ as a function of the G-band frequency $\omega_G$. A black dashed line indicates a slope of 2.2 corresponding to an evolution of the frequencies in the case of pure biaxial strain. The red dashed line shows the slope of 0.2 for the pure electron doping case. A pressure scale has been defined by sizing the markers proportionally to the pressure for each data point.}\label{Fig:2D:Correlation_2D-G}
\end{figure*}

\begin{figure*}
\centering
\includegraphics[width=.6\textwidth]{Plot_parameters_legend}

\smallskip

\begin{subfigure}{.49\textwidth}
\includegraphics[width = \textwidth]{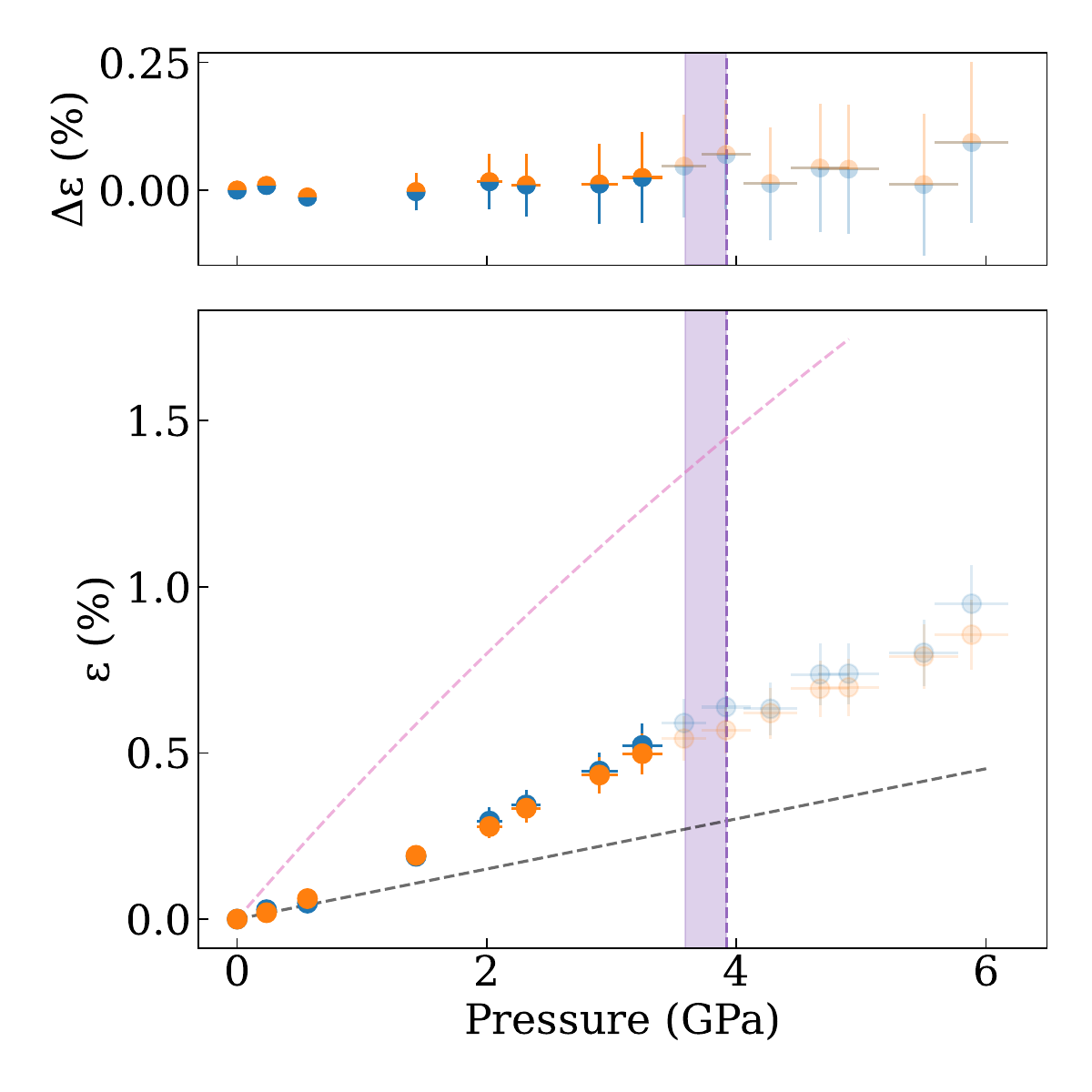} 
\caption{}\label{Fig:2D:strain}
\end{subfigure}
\begin{subfigure}{.49\textwidth}
\includegraphics[width = \textwidth]{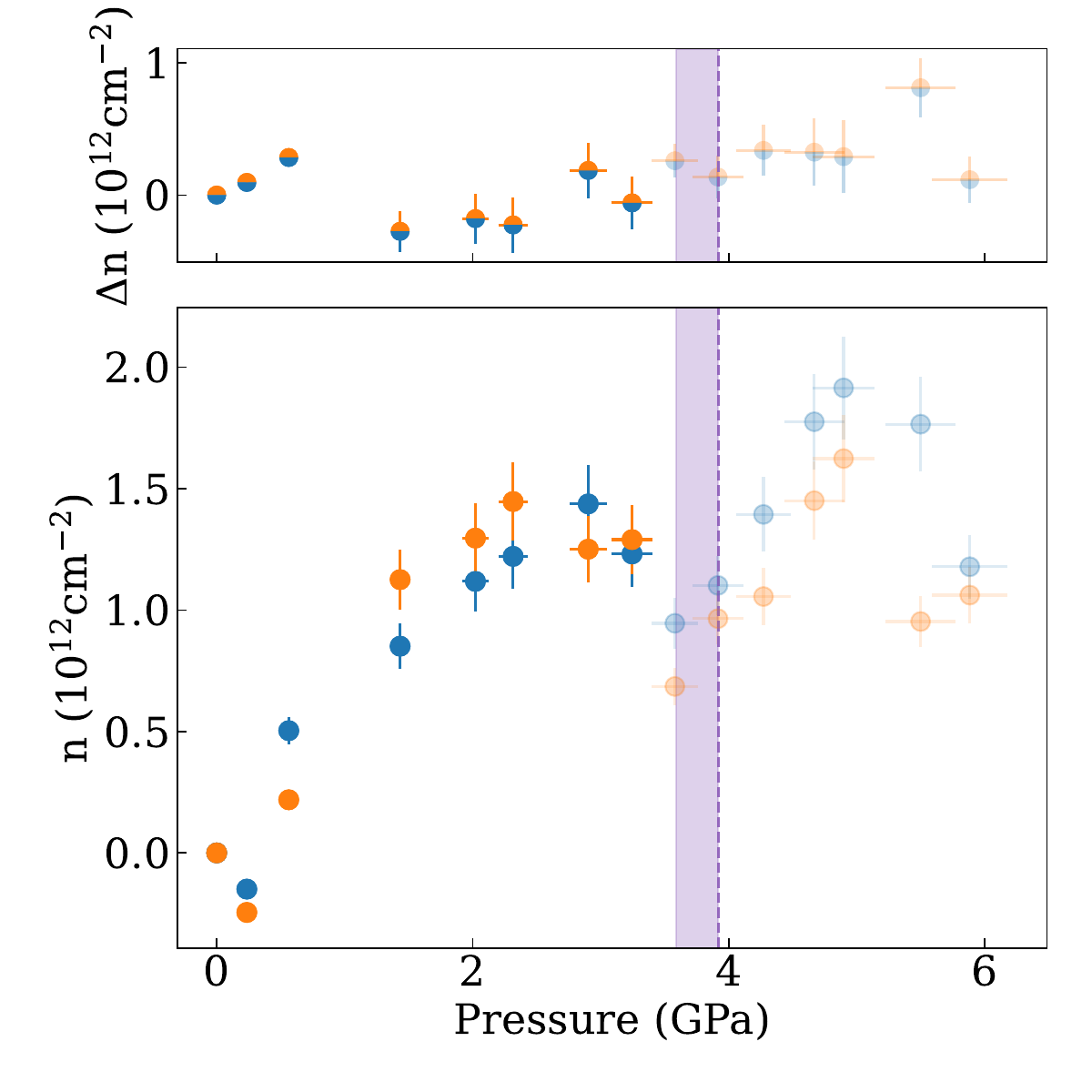} 
\caption{}\label{Fig:2D:doping}
\end{subfigure}%
\caption{(\subref{Fig:2D:strain}) Calculated biaxial strain $\epsilon$ and (\subref{Fig:2D:doping})  doping $n$ of the suspended and supported bilayer regions as a function of pressure. The values are calculated with respect to the point at ambient pressure for each region. The upper part of the plot shows the evolution of the difference in strain and doping between the suspended and the supported regions. In (\subref{Fig:2D:strain}), dashed lines are added to show two ideal cases: the strain evolution with pressure graphene would experience if it behaved like bulk graphite (black dashed line) and if it were in perfect adhesion with the substrate (pink dashed lines).}\label{Fig:2D:extracted}
\end{figure*}

The procedure proposed by Lee \textit{et al.} for disentangling the strain and doping contributions on graphene's Raman features evolution is based on the study of the correlation between $\omega_G$ and $\omega_{2D}$\cite{Lee2012}. The plotted values for our experiment are shown in \fig{Fig:2D:Correlation_2D-G}. Each marker has been sized to show the pressure at which the data was acquired. The slope $\partial\omega_{2D}/\partial\omega_{G}$ is found to follow the black dashed line in the case of pure strain applied to graphene. The latter corresponds to a slope $(\partial\omega_{2D}/\partial\omega_{G})^{strain} =2.2$\cite{Lee2012,Metten2014} crossing the suspended ambient pressure point. We note that most of the points lay in the region beneath. This evidence indicates that strain alone is not sufficient to describe the pressure evolution of the G and 2D-bands but doping must also be included. We consider here the case of electron doping of graphene from the PTM following our previous discussion on the evolution of $\Gamma_G$. The comparison with the hole doping scenario can be found in Figure~S3. While changing the absolute value of the strain and charge in the sample, our following considerations on the observed phenomena are valid in both cases. For pure electron doping we expect a slope  $(\partial\omega_{2D}/\partial\omega_{G})^{doping} = 0.2$\cite{Froehlicher2015}. We hence calculated the strain $\epsilon$ and the charge carrier density $n$ variations relative to the ambient pressure values\cite{Lee2012,Bendiab2018}. The evolution of the two quantities are shown in \fig{Fig:2D:extracted}.

For what concerns $\epsilon$ (\fig{Fig:2D:strain}), a monotonic, almost linear, increase with pressure is observed. The black dashed line represents the ideal behavior, assuming the graphene sample behaves like bulk graphite compressed under hydrostatic conditions\cite{Hanfland1989}. Our experiments show higher strain values throughout the pressure range. Indeed, when compressed at high pressure, studies indicate that the substrate induces an additional biaxial strain on supported graphene due to its volume reduction\cite{Bousige2017,vincent2024}. The ideal case of full adhesion and strain transmission from the substrate to the sample, calculated according to \citet{Decremps2010}, is shown by the pink dash line in \fig{Fig:2D:strain}. Our experiments show that the supported BLG attains intermediate strain values between the two ideal cases, in agreement with the evidence of partial biaxial strain transmission from the substrate to graphene previously reported\cite{Bousige2017}. Strikingly, however, no significant difference between the suspended and the supported regions is observed within our experimental resolution. This apparently counterintuitive result indicates that our suspended BLG acts as a rigid membrane, efficiently transferring the strain from the substrate to the suspended region.

The plot of the evolution of $n$ confirms the hypothesis of pressure induced doping effects in graphene peaking at values of $\sim1.4\cdot10^{12}$~\si{cm^{-2}}. Within the experimental uncertainties the doping levels are equivalent in the two regions, for the majority of the data points. A small additional doping of the suspended region is observed at low pressure, compatible with the doping effect induced by the PTM sandwiching the suspended BLG on two sides found in the previous section. Above 2~GPa, $n$ peaks to its maximum value retaining a seemingly constant behaviour until delamination of the substrate occurs. It is interesting to note that, similarly to what was observed in the evolution of $\Gamma_G$, we observe a decrease of the charge density at 0.2~GPa. This initial observation, considering that a charge injection of electrons and hole would result in a blue shift of the modes' frequencies\cite{Froehlicher2015}, is in agreement with our hypothesis of an inversion of the type of charge carrier from ambient condition to the pressurized system.

\subsection{Exploring local signatures by Raman spectral cartography}
\begin{figure*}
\begin{subfigure}{\textwidth}
\includegraphics[width = \textwidth]{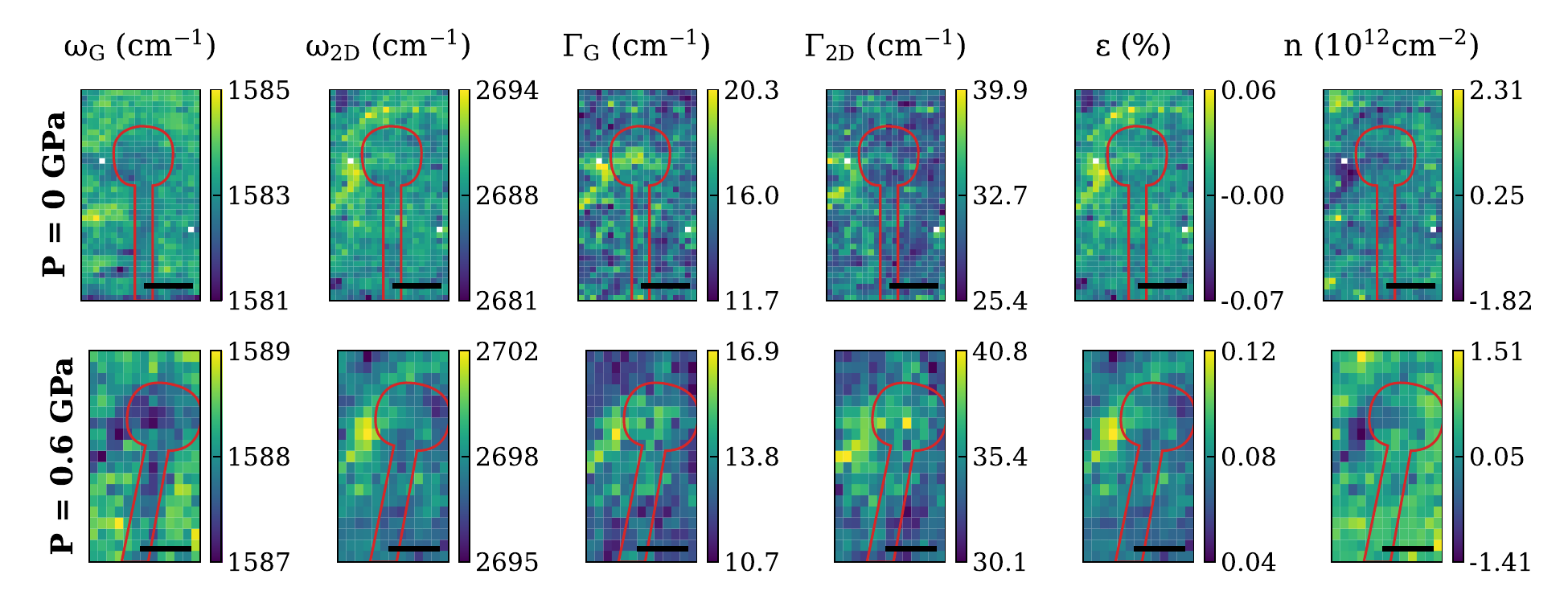}
\caption{}\label{Fig:Mapping}
\end{subfigure}

\begin{subfigure}{\textwidth}
\includegraphics[width = \textwidth]{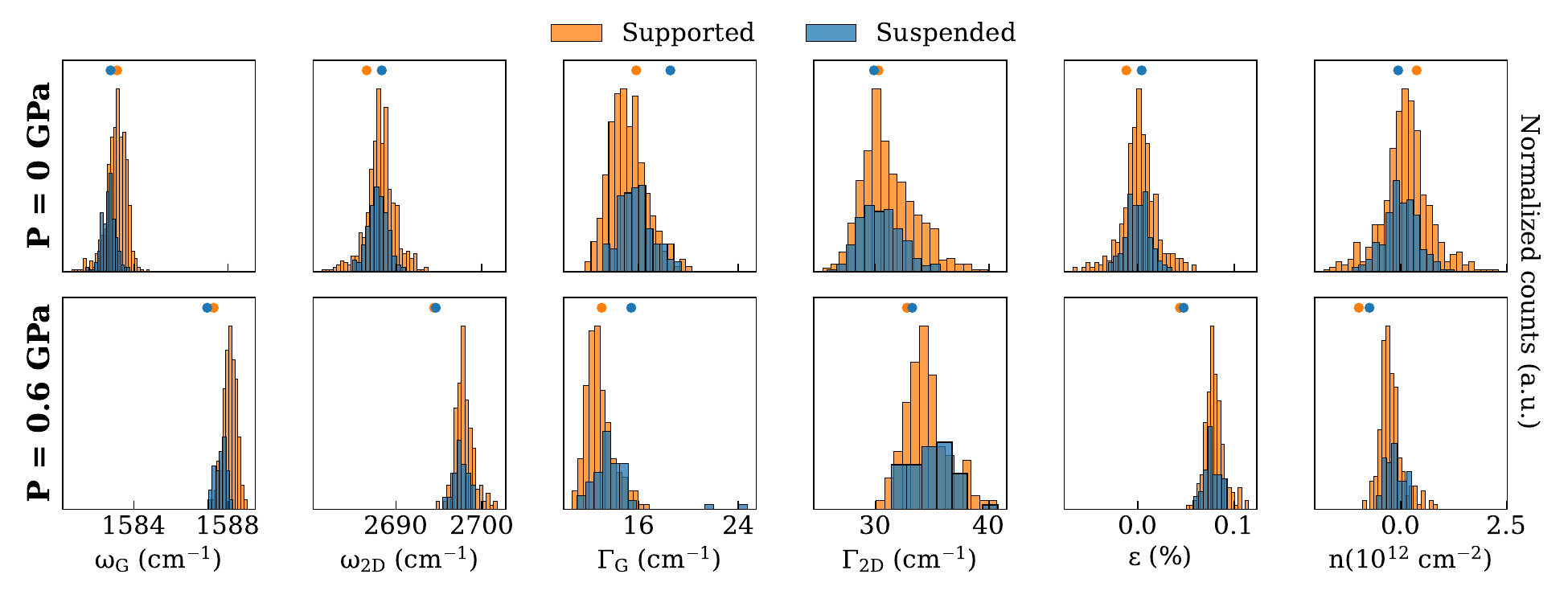} 
\caption{}\label{Fig:Mapping:Histos}
\end{subfigure}
\caption{(\subref{Fig:Mapping}) Raman spectral mapping of the sample at 0~GPa and 0.6~GPa. A black scale bar corresponding to 5~\si{\mu m} is shown on the bottom right of each plot. (\subref{Fig:Mapping:Histos}) Histograms of the parameters extracted from the maps in (\subref{Fig:Mapping}). Blue and orange dots are added above the histograms to locate the points at 0~\si{GPa} and 0.6~\si{GPa} that have been measured in the single point analysis for the suspended and supported regions respectively (\fig{Fig:2D:parameters} and \fig{Fig:2D:extracted}).}\label{}
\end{figure*}

\begin{table*}
\caption{Average values of the distributions in \fig{Fig:Mapping:Histos}. The uncertainties are given as one standard deviation of the distribution.}\label{Table:Mapping_average}
\begin{adjustbox}{width=1\textwidth,center}
\begin{tabular}{c c |c c c c c c c}
	&	&	$\omega_G$ (\si{cm^{-1}})&	$\omega_{2D}$ (\si{cm^{-1}})&	$\Gamma_G$ (\si{cm^{-1}})&	$\Gamma_{2D}$ (\si{cm^{-1}})&	$\epsilon$ (\%)&	$n $ (\si{10^{12} cm^{-2}})\\
	\hline
	\multirow{2}*{P=0 GPa} & Supported &	$1583.3\pm0.4$	&	$2688\pm2$	&	$15\pm2$	&	$31\pm2$	&	$0.00\pm0.02$	&	$0.1\pm0.6$	\\
		 & Suspended &	$1582.9\pm0.3$	&	$2688\pm1$	&	$16\pm1$	&	$31\pm2$	&	$-0.00\pm0.01$	&	$-0.0\pm0.4$	\\
	\multirow{2}*{P=0.6 GPa} & Supported &	$1588.1\pm0.3$	&	$2698\pm1$	&	$13\pm1$	&	$35\pm2$	&	$0.08\pm0.01$	&	$-0.2\pm0.3$	\\
		 & Suspended &	$1587.7\pm0.2$	&	$2697\pm1$	&	$14\pm1$	&	$35\pm2$	&	$0.08\pm0.01$	&	$-0.2\pm0.2$	\\
\end{tabular}
\end{adjustbox}
\end{table*}

The results we have obtained until now refer to two selected regions of the sample. Whilst providing a view on the evolution of the sample's properties throughout the pressure cycle, they lack a description of local variations in the pressure response of graphene's features. We, therefore, performed spatially resolved Raman spectral cartography of the sample in order to complement the previous results. We characterized the sample at ambient pressure and at 0.6~GPa, after the introduction of the PTM. A black rectangle in \fig{Fig:sample_P00} and \fig{Fig:sample_P02} highlights the regions that have been measured.

The fitted parameters for the G and 2D Raman peaks are displayed in the first four columns in \fig{Fig:Mapping}. A red line bounds the limits of the suspended region for a clearer visual inspection. The first row shows the spectra acquired at ambient pressure while the second one at 0.6~GPa. The data have than been processed in order to obtain the values of strain and doping with a similar procedure to that of the previous section. The values have been calculated in the present case with respect to the averaged value of the suspended region at ambient pressure. The strain and doping levels we obtained are plotted in the last two columns in \fig{Fig:Mapping}. In this case we calculated the doping values for holes at ambient pressure (substrate doping) and electrons at 0.6~GPa (PTM doping).

Local variations of the parameters are observed for all the spectra. A clear identification of the variations due to the sample in the supported and suspended region is not straightforward from the plots due to the inhomogeneities in the sample. We thus plotted the sample parameters into the histograms shown in \fig{Fig:Mapping:Histos}. The supported and suspended regions have been treated individually and they are represented by orange and blue histograms respectively. The identification of the two regions was possible thanks to the use of the diamond's Raman peak found around 1330~\cm{} at ambient pressure. It was present only in the suspended region at ambient pressure and with a considerably larger intensity in the same region during the compression cycle. The averaged values of the distributions are summarized in Table~\ref{Table:Mapping_average}. We find comparable values within the uncertainties for all the distributions. A systematic blue shift of the G-band in the suspended region is observed, however it lays within the standard deviation of the distributions. Moreover, we observe a decrease of $\Gamma_G$ with pressure, in agreement with our considerations of the previous section. 

Importantly, both strain and doping show very similar distributions for supported and suspended configurations confirming that a high degree of strain and doping transfer between the two regions is present when compressing the sample. Above each histogram we show two markers corresponding to the values of the parameters measured as single points discussed in the previous Section~\ref{Sec:Results}. This help us visualize where the studied regions are found with respect to the average distribution of the values across the sample. We observe that while single points do not perfectly match the quantitative average values of the parameters, they reflect well their overall evolution with pressure and between the two configurations. This illustrates the relevance of a spatially resolved mapping and call for further studies making use of it.

%% file: Chapters/Conclusion.tex
We reported on the spatially resolved study of the pressure response of a bilayer graphene sample which was partially suspended on a drilled substrate. 

Following the evolution of the Raman features at high pressure we identified the signatures of both doping and strain induced on our sample. To disentangle the individual effects of each contribution we plotted the evolution of the 2D-band frequency against that of the G-band. Our result shows that both strain and doping are not strongly affected by the suspended geometry, attending comparable values for the supported configuration within the studied pressure range and uncertainties. In the low pressure regime, higher value of charge transfer from the PTM are observed, indicating that the presence of the PTM on both sides of the sample can slightly enhance doping effects. Finally, the charge carrier density is shown to saturate fairly quickly within the first 2~GPa from where it becomes constant. We reached its maximum value of $(1.4\pm0.2)\cdot10^{12}$~\si{cm^{-2}} around this pressure.

The study was complemented by performing a spatially resolved Raman mapping at ambient pressure and at 0.6~GPa. To our knowledge this is the first case of diffraction-limit-resolved Raman cartography measurements at high pressure. This allowed us to identify the G-band as the mainly impacted feature by the suspension. The mean values of the distributions of each parameter has however revealed that, both at ambient pressure and at 0.6~GPa, the response of graphene in the two regions is comparable within one standard deviation. Moreover, in agreement with the results of Section~\ref{Sec:Disentangle}, both the average strain and doping values are comparable in the two regions. However, large local variations of the evolution of the Raman features emerge, revealing a much richer scenario than the usual \textit{single point} characterization of graphene experiments at high pressure. 

Those results motivate the possibility of novel approaches for the study of those systems. In particular, a detailed characterization through Raman mapping in the low pressure regime may reveal the presence of doping gradients which could open up opportunities for the construction of pressure tunable devices for energy harvesting\cite{HWANG2023} and electronics\cite{Gumprich2023,Wang2019b}. Finally, the significant strain transfer efficiency between the suspended and supported regions presents novel opportunities for studying samples in previously unattainable conditions necessitating strong biaxial strain whilst mitigating substrate effects and enhancing interaction with the environment.

%% file: main.bbl
\providecommand*{\mcitethebibliography}{\thebibliography}
\csname @ifundefined\endcsname{endmcitethebibliography}
{\let\endmcitethebibliography\endthebibliography}{}
\begin{mcitethebibliography}{42}
\providecommand*{\natexlab}[1]{#1}
\providecommand*{\mciteSetBstSublistMode}[1]{}
\providecommand*{\mciteSetBstMaxWidthForm}[2]{}
\providecommand*{\mciteBstWouldAddEndPuncttrue}
  {\def\EndOfBibitem{\unskip.}}
\providecommand*{\mciteBstWouldAddEndPunctfalse}
  {\let\EndOfBibitem\relax}
\providecommand*{\mciteSetBstMidEndSepPunct}[3]{}
\providecommand*{\mciteSetBstSublistLabelBeginEnd}[3]{}
\providecommand*{\EndOfBibitem}{}
\mciteSetBstSublistMode{f}
\mciteSetBstMaxWidthForm{subitem}
{(\emph{\alph{mcitesubitemcount}})}
\mciteSetBstSublistLabelBeginEnd{\mcitemaxwidthsubitemform\space}
{\relax}{\relax}

\bibitem[Novoselov \emph{et~al.}(2004)Novoselov, Geim, Morozov, Jiang, Zhang, Dubonos, Grigorieva, and Firsov]{Novoselov2004}
K.~S. Novoselov, A.~K. Geim, S.~V. Morozov, D.~Jiang, Y.~Zhang, S.~V. Dubonos, I.~V. Grigorieva and A.~A. Firsov, \emph{Science}, 2004, \textbf{306}, 666--669\relax
\mciteBstWouldAddEndPuncttrue
\mciteSetBstMidEndSepPunct{\mcitedefaultmidpunct}
{\mcitedefaultendpunct}{\mcitedefaultseppunct}\relax
\EndOfBibitem
\bibitem[Zhang \emph{et~al.}(2005)Zhang, Tan, Stormer, and Kim]{Zhang2005}
Y.~Zhang, Y.-W. Tan, H.~L. Stormer and P.~Kim, \emph{Nature}, 2005, \textbf{438}, 201--204\relax
\mciteBstWouldAddEndPuncttrue
\mciteSetBstMidEndSepPunct{\mcitedefaultmidpunct}
{\mcitedefaultendpunct}{\mcitedefaultseppunct}\relax
\EndOfBibitem
\bibitem[Yang \emph{et~al.}(2016)Yang, Wang, Wang, Xu, Tao, Zhang, Qin, Luther-Davies, Jagadish, Yu, and Lu]{Yang2016}
J.~Yang, Z.~Wang, F.~Wang, R.~Xu, J.~Tao, S.~Zhang, Q.~Qin, B.~Luther-Davies, C.~Jagadish, Z.~Yu and Y.~Lu, \emph{Light: Science {\&} Applications}, 2016, \textbf{5}, e16046--e16046\relax
\mciteBstWouldAddEndPuncttrue
\mciteSetBstMidEndSepPunct{\mcitedefaultmidpunct}
{\mcitedefaultendpunct}{\mcitedefaultseppunct}\relax
\EndOfBibitem
\bibitem[Wang \emph{et~al.}(2008)Wang, Zhang, Tian, Girit, Zettl, Crommie, and Shen]{Feng2008}
F.~Wang, Y.~Zhang, C.~Tian, C.~Girit, A.~Zettl, M.~Crommie and Y.~R. Shen, \emph{Science}, 2008, \textbf{320}, 206--209\relax
\mciteBstWouldAddEndPuncttrue
\mciteSetBstMidEndSepPunct{\mcitedefaultmidpunct}
{\mcitedefaultendpunct}{\mcitedefaultseppunct}\relax
\EndOfBibitem
\bibitem[Balandin \emph{et~al.}(2008)Balandin, Ghosh, Bao, Calizo, Teweldebrhan, Miao, and Lau]{Balandin2008}
A.~A. Balandin, S.~Ghosh, W.~Bao, I.~Calizo, D.~Teweldebrhan, F.~Miao and C.~N. Lau, \emph{Nano Letters}, 2008, \textbf{8}, 902--907\relax
\mciteBstWouldAddEndPuncttrue
\mciteSetBstMidEndSepPunct{\mcitedefaultmidpunct}
{\mcitedefaultendpunct}{\mcitedefaultseppunct}\relax
\EndOfBibitem
\bibitem[Lee \emph{et~al.}(2008)Lee, Wei, Kysar, and Hone]{Changgu2008}
C.~Lee, X.~Wei, J.~W. Kysar and J.~Hone, \emph{Science}, 2008, \textbf{321}, 385--388\relax
\mciteBstWouldAddEndPuncttrue
\mciteSetBstMidEndSepPunct{\mcitedefaultmidpunct}
{\mcitedefaultendpunct}{\mcitedefaultseppunct}\relax
\EndOfBibitem
\bibitem[Sun \emph{et~al.}(2021)Sun, Papageorgiou, Humphreys, Dunstan, Puech, Proctor, Bousige, Machon, and San-Miguel]{Sun2021}
Y.~W. Sun, D.~G. Papageorgiou, C.~J. Humphreys, D.~J. Dunstan, P.~Puech, J.~E. Proctor, C.~Bousige, D.~Machon and A.~San-Miguel, \emph{Applied Physics Reviews}, 2021, \textbf{8}, 021310\relax
\mciteBstWouldAddEndPuncttrue
\mciteSetBstMidEndSepPunct{\mcitedefaultmidpunct}
{\mcitedefaultendpunct}{\mcitedefaultseppunct}\relax
\EndOfBibitem
\bibitem[Bolotin \emph{et~al.}(2008)Bolotin, Sikes, Hone, Stormer, and Kim]{Bolotin2008}
K.~I. Bolotin, K.~J. Sikes, J.~Hone, H.~L. Stormer and P.~Kim, \emph{Phys. Rev. Lett.}, 2008, \textbf{101}, 096802\relax
\mciteBstWouldAddEndPuncttrue
\mciteSetBstMidEndSepPunct{\mcitedefaultmidpunct}
{\mcitedefaultendpunct}{\mcitedefaultseppunct}\relax
\EndOfBibitem
\bibitem[Taube \emph{et~al.}(2015)Taube, Judek, Łapińska, and Zdrojek]{Taube2015}
A.~Taube, J.~Judek, A.~Łapińska and M.~Zdrojek, \emph{ACS Applied Materials \& Interfaces}, 2015, \textbf{7}, 5061--5065\relax
\mciteBstWouldAddEndPuncttrue
\mciteSetBstMidEndSepPunct{\mcitedefaultmidpunct}
{\mcitedefaultendpunct}{\mcitedefaultseppunct}\relax
\EndOfBibitem
\bibitem[Burch \emph{et~al.}(2018)Burch, Mandrus, and Park]{Burch2018}
K.~S. Burch, D.~Mandrus and J.-G. Park, \emph{Nature}, 2018, \textbf{563}, 47--52\relax
\mciteBstWouldAddEndPuncttrue
\mciteSetBstMidEndSepPunct{\mcitedefaultmidpunct}
{\mcitedefaultendpunct}{\mcitedefaultseppunct}\relax
\EndOfBibitem
\bibitem[Mak \emph{et~al.}(2019)Mak, Shan, and Ralph]{Mak2019}
K.~F. Mak, J.~Shan and D.~C. Ralph, \emph{Nature Reviews Physics}, 2019, \textbf{1}, 646--661\relax
\mciteBstWouldAddEndPuncttrue
\mciteSetBstMidEndSepPunct{\mcitedefaultmidpunct}
{\mcitedefaultendpunct}{\mcitedefaultseppunct}\relax
\EndOfBibitem
\bibitem[Zhang \emph{et~al.}(2009)Zhang, Tang, Girit, Hao, Martin, Zettl, Crommie, Shen, and Wang]{Zhang2009}
Y.~Zhang, T.-T. Tang, C.~Girit, Z.~Hao, M.~C. Martin, A.~Zettl, M.~F. Crommie, Y.~R. Shen and F.~Wang, \emph{Nature}, 2009, \textbf{459}, 820--823\relax
\mciteBstWouldAddEndPuncttrue
\mciteSetBstMidEndSepPunct{\mcitedefaultmidpunct}
{\mcitedefaultendpunct}{\mcitedefaultseppunct}\relax
\EndOfBibitem
\bibitem[Jauregui \emph{et~al.}(2019)Jauregui, Joe, Pistunova, Wild, High, Zhou, Scuri, Greve, Sushko, Yu, Taniguchi, Watanabe, Needleman, Lukin, Park, and Kim]{Jautegui2019}
L.~A. Jauregui, A.~Y. Joe, K.~Pistunova, D.~S. Wild, A.~A. High, Y.~Zhou, G.~Scuri, K.~D. Greve, A.~Sushko, C.-H. Yu, T.~Taniguchi, K.~Watanabe, D.~J. Needleman, M.~D. Lukin, H.~Park and P.~Kim, \emph{Science}, 2019, \textbf{366}, 870--875\relax
\mciteBstWouldAddEndPuncttrue
\mciteSetBstMidEndSepPunct{\mcitedefaultmidpunct}
{\mcitedefaultendpunct}{\mcitedefaultseppunct}\relax
\EndOfBibitem
\bibitem[Chiout \emph{et~al.}(2023)Chiout, Brochard-Richard, Marty, Bendiab, Zhao, Johnson, Oehler, Ouerghi, and Chaste]{Chiout2023}
A.~Chiout, C.~Brochard-Richard, L.~Marty, N.~Bendiab, M.-Q. Zhao, A.~T.~C. Johnson, F.~Oehler, A.~Ouerghi and J.~Chaste, \emph{npj 2D Materials and Applications}, 2023, \textbf{7}, 20\relax
\mciteBstWouldAddEndPuncttrue
\mciteSetBstMidEndSepPunct{\mcitedefaultmidpunct}
{\mcitedefaultendpunct}{\mcitedefaultseppunct}\relax
\EndOfBibitem
\bibitem[Medeghini \emph{et~al.}(2018)Medeghini, Hettich, Rouxel, Silva~Santos, Hermelin, Pertreux, Torres~Dias, Legrand, Maioli, Crut, Vallée, San~Miguel, and Del~Fatti]{Medeghini2018}
F.~Medeghini, M.~Hettich, R.~Rouxel, S.~D. Silva~Santos, S.~Hermelin, E.~Pertreux, A.~Torres~Dias, F.~Legrand, P.~Maioli, A.~Crut, F.~Vallée, A.~San~Miguel and N.~Del~Fatti, \emph{ACS Nano}, 2018, \textbf{12}, 10310--10316\relax
\mciteBstWouldAddEndPuncttrue
\mciteSetBstMidEndSepPunct{\mcitedefaultmidpunct}
{\mcitedefaultendpunct}{\mcitedefaultseppunct}\relax
\EndOfBibitem
\bibitem[Machon \emph{et~al.}(2018)Machon, Pischedda, Le~Floch, and San-Miguel]{Machon2018b}
D.~Machon, V.~Pischedda, S.~Le~Floch and A.~San-Miguel, \emph{Journal of Applied Physics}, 2018, \textbf{124}, 160902\relax
\mciteBstWouldAddEndPuncttrue
\mciteSetBstMidEndSepPunct{\mcitedefaultmidpunct}
{\mcitedefaultendpunct}{\mcitedefaultseppunct}\relax
\EndOfBibitem
\bibitem[Martins \emph{et~al.}(2017)Martins, Matos, Paschoal, Freire, Andrade, Aguiar, Kong, Neves, de~Oliveira, Mazzoni, Filho, and Can{\c{c}}ado]{Martins2017}
L.~G.~P. Martins, M.~J.~S. Matos, A.~R. Paschoal, P.~T.~C. Freire, N.~F. Andrade, A.~L. Aguiar, J.~Kong, B.~R.~A. Neves, A.~B. de~Oliveira, M.~S. Mazzoni, A.~G.~S. Filho and L.~G. Can{\c{c}}ado, \emph{Nature Communications}, 2017, \textbf{8}, 96\relax
\mciteBstWouldAddEndPuncttrue
\mciteSetBstMidEndSepPunct{\mcitedefaultmidpunct}
{\mcitedefaultendpunct}{\mcitedefaultseppunct}\relax
\EndOfBibitem
\bibitem[Vincent \emph{et~al.}(2024)Vincent, Galafassi, Hellani, Forestier, Brigiano, Araujo, Piednoir, Diaf, Pietrucci, Filho, del Fatti, Vialla, and San-Miguel]{vincent2024}
R.~Vincent, R.~Galafassi, M.~Hellani, A.~Forestier, F.~S. Brigiano, B.~S. Araujo, A.~Piednoir, H.~Diaf, F.~Pietrucci, A.~G.~S. Filho, N.~del Fatti, F.~Vialla and A.~San-Miguel, \emph{Biaxial strain effects in 2D diamond formation from graphene stacks}, 2024, \url{https://arxiv.org/abs/2405.06416}\relax
\mciteBstWouldAddEndPuncttrue
\mciteSetBstMidEndSepPunct{\mcitedefaultmidpunct}
{\mcitedefaultendpunct}{\mcitedefaultseppunct}\relax
\EndOfBibitem
\bibitem[Bronsgeest \emph{et~al.}(2015)Bronsgeest, Bendiab, Mathur, Kimouche, Johnson, Coraux, and Pochet]{Bronsgeest2015}
M.~S. Bronsgeest, N.~Bendiab, S.~Mathur, A.~Kimouche, H.~T. Johnson, J.~Coraux and P.~Pochet, \emph{Nano Letters}, 2015, \textbf{15}, 5098--5104\relax
\mciteBstWouldAddEndPuncttrue
\mciteSetBstMidEndSepPunct{\mcitedefaultmidpunct}
{\mcitedefaultendpunct}{\mcitedefaultseppunct}\relax
\EndOfBibitem
\bibitem[Schmidt \emph{et~al.}(2011)Schmidt, Ohta, and Beechem]{Schmidt2011}
D.~A. Schmidt, T.~Ohta and T.~E. Beechem, \emph{Phys. Rev. B}, 2011, \textbf{84}, 235422\relax
\mciteBstWouldAddEndPuncttrue
\mciteSetBstMidEndSepPunct{\mcitedefaultmidpunct}
{\mcitedefaultendpunct}{\mcitedefaultseppunct}\relax
\EndOfBibitem
\bibitem[Fan(2011)]{Fan2011}
\emph{The Journal of Physical Chemistry C}, 2011, \textbf{115}, 12960--12964\relax
\mciteBstWouldAddEndPuncttrue
\mciteSetBstMidEndSepPunct{\mcitedefaultmidpunct}
{\mcitedefaultendpunct}{\mcitedefaultseppunct}\relax
\EndOfBibitem
\bibitem[Ji \emph{et~al.}(2021)Ji, Kim, Lee, Sung, Kim, Park, Hong, and Lee]{JI2021}
E.~Ji, M.~J. Kim, J.-Y. Lee, D.~Sung, N.~Kim, J.-W. Park, S.~Hong and G.-H. Lee, \emph{Carbon}, 2021, \textbf{184}, 651--658\relax
\mciteBstWouldAddEndPuncttrue
\mciteSetBstMidEndSepPunct{\mcitedefaultmidpunct}
{\mcitedefaultendpunct}{\mcitedefaultseppunct}\relax
\EndOfBibitem
\bibitem[Sun \emph{et~al.}(2022)Sun, Holec, Gehringer, Li, Fenwick, Dunstan, and Humphreys]{Sun2022}
Y.~W. Sun, D.~Holec, D.~Gehringer, L.~Li, O.~Fenwick, D.~J. Dunstan and C.~J. Humphreys, \emph{Phys. Rev. B}, 2022, \textbf{105}, 165416\relax
\mciteBstWouldAddEndPuncttrue
\mciteSetBstMidEndSepPunct{\mcitedefaultmidpunct}
{\mcitedefaultendpunct}{\mcitedefaultseppunct}\relax
\EndOfBibitem
\bibitem[Proctor \emph{et~al.}(2006)Proctor, Halsall, Ghandour, and Dunstan]{Proctor2006}
J.~E. Proctor, M.~P. Halsall, A.~Ghandour and D.~J. Dunstan, \emph{Journal of Physics and Chemistry of Solids}, 2006, \textbf{67}, 2468--2472\relax
\mciteBstWouldAddEndPuncttrue
\mciteSetBstMidEndSepPunct{\mcitedefaultmidpunct}
{\mcitedefaultendpunct}{\mcitedefaultseppunct}\relax
\EndOfBibitem
\bibitem[Machon \emph{et~al.}(2018)Machon, Bousige, Alencar, Torres-Dias, Balima, Nicolle, de~Sousa~Pinheiro, Souza~Filho, and San-Miguel]{Machon2018}
D.~Machon, C.~Bousige, R.~Alencar, A.~Torres-Dias, F.~Balima, J.~Nicolle, G.~de~Sousa~Pinheiro, A.~G. Souza~Filho and A.~San-Miguel, \emph{Journal of Raman Spectroscopy}, 2018, \textbf{49}, 121--129\relax
\mciteBstWouldAddEndPuncttrue
\mciteSetBstMidEndSepPunct{\mcitedefaultmidpunct}
{\mcitedefaultendpunct}{\mcitedefaultseppunct}\relax
\EndOfBibitem
\bibitem[Forestier \emph{et~al.}(2020)Forestier, Balima, Bousige, Pinheiro, Fulcrand, Kalbáč, Machon, and San-Miguel]{Forestier2020}
A.~Forestier, F.~Balima, C.~Bousige, G.~d.~S. Pinheiro, R.~Fulcrand, M.~Kalbáč, D.~Machon and A.~San-Miguel, \emph{The Journal of Physical Chemistry C}, 2020, \textbf{124}, 11193--11199\relax
\mciteBstWouldAddEndPuncttrue
\mciteSetBstMidEndSepPunct{\mcitedefaultmidpunct}
{\mcitedefaultendpunct}{\mcitedefaultseppunct}\relax
\EndOfBibitem
\bibitem[Bousige \emph{et~al.}(2017)Bousige, Balima, Machon, Pinheiro, Torres-Dias, Nicolle, Kalita, Bendiab, Marty, Bouchiat, Montagnac, Souza~Filho, Poncharal, and San-Miguel]{Bousige2017}
C.~Bousige, F.~Balima, D.~Machon, G.~S. Pinheiro, A.~Torres-Dias, J.~Nicolle, D.~Kalita, N.~Bendiab, L.~Marty, V.~Bouchiat, G.~Montagnac, A.~G. Souza~Filho, P.~Poncharal and A.~San-Miguel, \emph{Nano Letters}, 2017, \textbf{17}, 21--27\relax
\mciteBstWouldAddEndPuncttrue
\mciteSetBstMidEndSepPunct{\mcitedefaultmidpunct}
{\mcitedefaultendpunct}{\mcitedefaultseppunct}\relax
\EndOfBibitem
\bibitem[Lee \emph{et~al.}(2012)Lee, Ahn, Shim, Lee, and Ryu]{Lee2012}
J.~E. Lee, G.~Ahn, J.~Shim, Y.~S. Lee and S.~Ryu, \emph{Nature Communications}, 2012, \textbf{3}, 1024\relax
\mciteBstWouldAddEndPuncttrue
\mciteSetBstMidEndSepPunct{\mcitedefaultmidpunct}
{\mcitedefaultendpunct}{\mcitedefaultseppunct}\relax
\EndOfBibitem
\bibitem[Klotz \emph{et~al.}(2009)Klotz, Chervin, Munsch, and Marchand]{Klotz2009}
S.~Klotz, J.-C. Chervin, P.~Munsch and G.~L. Marchand, \emph{Journal of Physics D: Applied Physics}, 2009, \textbf{42}, 075413\relax
\mciteBstWouldAddEndPuncttrue
\mciteSetBstMidEndSepPunct{\mcitedefaultmidpunct}
{\mcitedefaultendpunct}{\mcitedefaultseppunct}\relax
\EndOfBibitem
\bibitem[Nicolle \emph{et~al.}(2011)Nicolle, Machon, Poncharal, Pierre-Louis, and San-Miguel]{Nicolle2011}
J.~Nicolle, D.~Machon, P.~Poncharal, O.~Pierre-Louis and A.~San-Miguel, \emph{Nano Letters}, 2011, \textbf{11}, 3564--3568\relax
\mciteBstWouldAddEndPuncttrue
\mciteSetBstMidEndSepPunct{\mcitedefaultmidpunct}
{\mcitedefaultendpunct}{\mcitedefaultseppunct}\relax
\EndOfBibitem
\bibitem[Chijioke \emph{et~al.}(2005)Chijioke, Nellis, Soldatov, and Silvera]{Chijioke2005}
A.~D. Chijioke, W.~J. Nellis, A.~Soldatov and I.~F. Silvera, \emph{Journal of Applied Physics}, 2005, \textbf{98}, 114905\relax
\mciteBstWouldAddEndPuncttrue
\mciteSetBstMidEndSepPunct{\mcitedefaultmidpunct}
{\mcitedefaultendpunct}{\mcitedefaultseppunct}\relax
\EndOfBibitem
\bibitem[Amaral \emph{et~al.}(2021)Amaral, Forestier, Piednoir, Galafassi, Bousige, Machon, Pierre-Louis, Alencar, {Souza Filho}, and San-Miguel]{AMARAL2021}
I.~Amaral, A.~Forestier, A.~Piednoir, R.~Galafassi, C.~Bousige, D.~Machon, O.~Pierre-Louis, R.~Alencar, A.~{Souza Filho} and A.~San-Miguel, \emph{Carbon}, 2021, \textbf{185}, 242--251\relax
\mciteBstWouldAddEndPuncttrue
\mciteSetBstMidEndSepPunct{\mcitedefaultmidpunct}
{\mcitedefaultendpunct}{\mcitedefaultseppunct}\relax
\EndOfBibitem
\bibitem[Froehlicher and Berciaud(2015)]{Froehlicher2015}
G.~Froehlicher and S.~Berciaud, \emph{Phys. Rev. B}, 2015, \textbf{91}, 205413\relax
\mciteBstWouldAddEndPuncttrue
\mciteSetBstMidEndSepPunct{\mcitedefaultmidpunct}
{\mcitedefaultendpunct}{\mcitedefaultseppunct}\relax
\EndOfBibitem
\bibitem[Aguirre \emph{et~al.}(2009)Aguirre, Levesque, Paillet, Lapointe, St-Antoine, Desjardins, and Martel]{Aguirre2009}
C.~M. Aguirre, P.~L. Levesque, M.~Paillet, F.~Lapointe, B.~C. St-Antoine, P.~Desjardins and R.~Martel, \emph{Advanced Materials}, 2009, \textbf{21}, 3087--3091\relax
\mciteBstWouldAddEndPuncttrue
\mciteSetBstMidEndSepPunct{\mcitedefaultmidpunct}
{\mcitedefaultendpunct}{\mcitedefaultseppunct}\relax
\EndOfBibitem
\bibitem[Wilder \emph{et~al.}(1998)Wilder, Venema, Rinzler, Smalley, and Dekker]{Wilder1998}
J.~W.~G. Wilder, L.~C. Venema, A.~G. Rinzler, R.~E. Smalley and C.~Dekker, \emph{Nature}, 1998, \textbf{391}, 59--62\relax
\mciteBstWouldAddEndPuncttrue
\mciteSetBstMidEndSepPunct{\mcitedefaultmidpunct}
{\mcitedefaultendpunct}{\mcitedefaultseppunct}\relax
\EndOfBibitem
\bibitem[Metten \emph{et~al.}(2014)Metten, Federspiel, Romeo, and Berciaud]{Metten2014}
D.~Metten, F.~m.~c. Federspiel, M.~Romeo and S.~Berciaud, \emph{Phys. Rev. Appl.}, 2014, \textbf{2}, 054008\relax
\mciteBstWouldAddEndPuncttrue
\mciteSetBstMidEndSepPunct{\mcitedefaultmidpunct}
{\mcitedefaultendpunct}{\mcitedefaultseppunct}\relax
\EndOfBibitem
\bibitem[Bendiab \emph{et~al.}(2018)Bendiab, Renard, Schwarz, Reserbat-Plantey, Djevahirdjian, Bouchiat, Coraux, and Marty]{Bendiab2018}
N.~Bendiab, J.~Renard, C.~Schwarz, A.~Reserbat-Plantey, L.~Djevahirdjian, V.~Bouchiat, J.~Coraux and L.~Marty, \emph{Journal of Raman Spectroscopy}, 2018, \textbf{49}, 130--145\relax
\mciteBstWouldAddEndPuncttrue
\mciteSetBstMidEndSepPunct{\mcitedefaultmidpunct}
{\mcitedefaultendpunct}{\mcitedefaultseppunct}\relax
\EndOfBibitem
\bibitem[Hanfland \emph{et~al.}(1989)Hanfland, Beister, and Syassen]{Hanfland1989}
M.~Hanfland, H.~Beister and K.~Syassen, \emph{Phys. Rev. B}, 1989, \textbf{39}, 12598--12603\relax
\mciteBstWouldAddEndPuncttrue
\mciteSetBstMidEndSepPunct{\mcitedefaultmidpunct}
{\mcitedefaultendpunct}{\mcitedefaultseppunct}\relax
\EndOfBibitem
\bibitem[Decremps \emph{et~al.}(2010)Decremps, Belliard, Gauthier, and Perrin]{Decremps2010}
F.~Decremps, L.~Belliard, M.~Gauthier and B.~Perrin, \emph{Phys. Rev. B}, 2010, \textbf{82}, 104119\relax
\mciteBstWouldAddEndPuncttrue
\mciteSetBstMidEndSepPunct{\mcitedefaultmidpunct}
{\mcitedefaultendpunct}{\mcitedefaultseppunct}\relax
\EndOfBibitem
\bibitem[Hwang \emph{et~al.}(2023)Hwang, Kim, Lee, and Lee]{HWANG2023}
H.~J. Hwang, S.-Y. Kim, S.~K. Lee and B.~H. Lee, \emph{Carbon}, 2023, \textbf{201}, 467--472\relax
\mciteBstWouldAddEndPuncttrue
\mciteSetBstMidEndSepPunct{\mcitedefaultmidpunct}
{\mcitedefaultendpunct}{\mcitedefaultseppunct}\relax
\EndOfBibitem
\bibitem[Gumprich \emph{et~al.}(2023)Gumprich, Liedtke, Beck, Chirca, Potočnik, Alexander-Webber, Hofmann, and Tappertzhofen]{Gumprich2023}
A.~Gumprich, J.~Liedtke, S.~Beck, I.~Chirca, T.~Potočnik, J.~A. Alexander-Webber, S.~Hofmann and S.~Tappertzhofen, \emph{Nanotechnology}, 2023, \textbf{34}, 265203\relax
\mciteBstWouldAddEndPuncttrue
\mciteSetBstMidEndSepPunct{\mcitedefaultmidpunct}
{\mcitedefaultendpunct}{\mcitedefaultseppunct}\relax
\EndOfBibitem
\bibitem[Wang \emph{et~al.}(2019)Wang, Liu, Hao, Wang, Chen, Li, and Dong]{Wang2019b}
Z.~Wang, J.~Liu, X.~Hao, Y.~Wang, Y.~Chen, P.~Li and M.~Dong, \emph{New J. Chem.}, 2019, \textbf{43}, 15275--15279\relax
\mciteBstWouldAddEndPuncttrue
\mciteSetBstMidEndSepPunct{\mcitedefaultmidpunct}
{\mcitedefaultendpunct}{\mcitedefaultseppunct}\relax
\EndOfBibitem
\end{mcitethebibliography}
